%% file: main.tex
\documentclass{article}
\usepackage{iclr2027_conference,times}

\input{math_commands.tex}

\usepackage{hyperref}
\usepackage{url}
\usepackage{booktabs}
\usepackage{multirow}
\usepackage{graphicx}
\usepackage{xcolor}
\usepackage{wrapfig}
\usepackage{tikz}
\usetikzlibrary{positioning,arrows.meta,calc,fit,decorations.pathreplacing}

\title{WHEN DIFFUSION RECOMMENDERS STOP SAMPLING: PRESERVING GENERATIVE DIVERSITY}

\author{%
{\footnotesize
Louis Castres Saint Martin$^{1,2}$,
Alain Rakotomamonjy$^{1}$,
Patrick Gallinari$^{1,2}$}\\[0.75em]
$^{1}$Criteo AI Lab, Paris, France \\
$^{2}$Sorbonne Universit\'e, CNRS, ISIR, 75005 Paris, France
}

\iclrfinalcopy

\begin{document}

\maketitle
\lhead{Under review as a conference paper at ICLR 2027}
\makeatother

\input{content/abstract.tex}
\input{content/introduction.tex}
\input{content/related-work.tex}
\input{content/method.tex}
\input{content/experiments.tex}
\input{content/discussion.tex}
\input{appendix/ai-use-statement}
\input{content/acknowledgment.tex}

\bibliographystyle{iclr2027_conference}
\bibliography{main}

\appendix
\input{appendix/implementation-details.tex}

\input{appendix/item-representation-ablation}
\input{appendix/ce-ablation.tex}
\input{appendix/design-ablation.tex}
\input{appendix/stochastic-sasrec.tex}
\input{appendix/union-recall.tex}
\input{appendix/inference-time}

\end{document}

%% file: math_commands.tex
\usepackage{amsmath,amsfonts,bm}

\def\eqref#1{equation~\ref{#1}}

\def\1{\bm{1}}

\def\vzero{{\bm{0}}}

\def\vc{{\bm{c}}}

\def\ve{{\bm{e}}}

\def\vh{{\bm{h}}}

\def\vl{{\bm{l}}}

\def\vz{{\bm{z}}}

\def\mC{{\bm{C}}}

\def\mI{{\bm{I}}}

\DeclareMathAlphabet{\mathsfit}{\encodingdefault}{\sfdefault}{m}{sl}
\SetMathAlphabet{\mathsfit}{bold}{\encodingdefault}{\sfdefault}{bx}{n}

\newcommand{\E}{\mathbb{E}}



%% file: content/abstract.tex
\begin{abstract}
    A user history can admit multiple plausible next interactions, yet conventional sequential recommenders produce a single deterministic ranking. While diffusion models offer a natural way to sample alternative continuations, we show that existing target-item diffusion recommenders often suffer from severe sampling collapse, producing nearly identical outputs across independent draws. We identify two ubiquitous design choices that promote this collapse: joint learning of the latent space and denoiser, and in-trajectory supervision via discrete losses. To address these failure modes, we introduce \textbf{DARTS} (\textbf{D}iverse and \textbf{A}ccurate \textbf{R}ecommendation \textbf{T}hrough \textbf{S}ampling), a three-stage latent diffusion framework that explicitly decouples representation learning, generative modeling, and discrete ranking by learning and freezing the latent space before diffusion training and postponing ranking supervision to the final decoding stage. We further characterize sampling diversity using complementary measures of variation before and after decoding, and use posterior-predictive aggregation to exploit this diversity at inference. Across three Amazon Reviews 2023 categories, DARTS preserves substantially greater sampling diversity while outperforming strong sequential, generative-retrieval, and diffusion baselines by 7--12\% on standard ranking metrics. Aggregating multiple samples further improves ranking accuracy at a fixed slate size.
    \end{abstract}

%% file: content/introduction.tex
\section{Introduction}
\label{sec:introduction}



Sequential recommendation aims to predict the conditional distribution of users' future interactions given their past behavior. Most commonly, these systems are trained to predict the next item given the preceding interactions. Given the diversity of user behaviors and the large size of the item space, these conditional distributions can be difficult to model accurately.

The dominant approach relies on Transformer-based recommenders, from SASRec~\citep{kang2018sasrec} and BERT4Rec~\citep{sun2019bert4rec} to their extensions to industrial-scale catalogues~\citep{zhai2024hstu} and semantic identifiers~\citep{rajput2023tiger}. Despite architectural differences, they share one training paradigm: the history is encoded into a single representation, scored against the catalogue, and optimised with a catalogue cross-entropy or ranking loss. Although effective, such discriminative models may struggle to capture the diversity of user behaviour. Their output distribution can in principle be sampled to obtain varied predictions, but it does not faithfully reflect that diversity, and sampling from it degrades accuracy (Figure~\ref{fig:stochastic-sasrec}). These models are therefore used deterministically, returning the same ranking for the same history.

Diffusion models~\citep{ho2020ddpm,song2021sde} provide an alternative to discriminative models for learning complex distributions, and a growing body of work has recently incorporated them into recommender systems~\citep{wang2023diffrec,zhao2024ddrm,li2023diffurec,yang2023dreamrec,chen2025adrec,liu2025preferdiff,hu2025prefergrow}. Because generation starts from random noise, two samples conditioned on the same user history can, in principle, lead to different items. This gives diffusion-based recommenders the potential to preserve multiple plausible continuations rather than collapsing them into a single prediction~\citep{li2023diffurec,zolghadr2024diffurecsys}.
In practice, however, we find that this diversity largely fails to materialize. In most settings, repeatedly sampling existing diffusion recommenders for the same user history yields nearly identical rankings and only exhibit a weak diversity (Section~\ref{ssec:generativeness}), so that the resulting sampler behaves almost deterministically.

We attribute this behavior to two design choices. First, item embeddings are learned jointly with the denoiser, allowing the latent space to reshape itself until little uncertainty remains for the denoising process to model. Second, the denoiser's intermediate outputs are supervised using catalogue-level cross-entropy or ranking losses, which encourage different denoising trajectories to converge towards the same high-scoring items. Said otherwise, relevance based recommenders naturally produce homogeneous and redundant lists. In this setting, we observe a tension between diversity and recommendation performance, as measured by accuracy metrics such as Hit Rate and NDCG: increasing diversity typically comes at the expense of recommendation accuracy, and vice versa. This is known as the diversity-accuracy dilemma \cite{peng2024reconciling}.
Successful continuous generative models avoid these design choices. Latent diffusion models for images fix the representation before training the generative model~\citep{rombach2022ldm}, while recent continuous language models such as ELF learn the mapping back to discrete tokens only at the final stage~\citep{hu2026elf}.

We carry these two principles over to sequential recommendation and introduce \textbf{DARTS} (\textbf{D}iverse and \textbf{A}ccurate \textbf{R}ecommendation \textbf{T}hrough \textbf{S}ampling), a generative recommendation model that can simultaneously improve prediction diversity and recommendation quality relative to the discriminative and diffusion-based baselines considered in our experiments.
To the best of our knowledge, DARTS is among the first generative sequential recommenders to improve both sampling diversity and ranking performance, suggesting that generative modeling can capture multiple plausible user continuations without sacrificing recommendation quality. DARTS proceeds in three stages. We first pretrain a sequence autoencoder and freeze its latent space. We then train a denoiser in this fixed space using only a regression loss, without catalogue-level supervision. Finally, we fine-tune the decoder with a ranking objective. Sampling thus takes place entirely in the continuous latent space, allowing repeated draws to remain diverse.

Our contributions are the following.
\begin{itemize}
  \item DARTS, a three-stage latent diffusion model for sequential recommendation that separates representation learning, generative training, and catalogue decoding (Section~\ref{sec:method}).
  \item Sampling diagnostics, complementary to ranking metrics, that measure the diversity of repeated draws in latent and item space. With them, we show that existing diffusion recommenders collapse to near-deterministic sampling, and we trace this collapse to joint latent training and in-trajectory catalogue supervision.
  \item Experiments on three categories of Amazon Reviews 2023~\citep{hou2024amazon2023}, where DARTS improves over discriminative, generative-retrieval, and diffusion baselines by 6.8--12.0\% on standard ranking metrics. Aggregating several posterior samples into a single slate of fixed size gives further gains.
\end{itemize}

%% file: content/related-work.tex
\section{Related Work}
\label{sec:related-work}

\paragraph{Sequential recommendation.}
GRU4Rec~\citep{hidasi2016gru4rec} applied recurrent networks to interaction sequences, and later models moved to transformer encoders.
SASRec~\citep{kang2018sasrec} is trained with causal next-item prediction, BERT4Rec~\citep{sun2019bert4rec} with masked item modelling, and S\textsuperscript{3}-Rec~\citep{zhou2020s3rec} adds further self-supervised objectives.
More recent work scales this recipe to industrial catalogues~\citep{zhai2024hstu}, or retrieves items by generating their semantic identifiers autoregressively~\citep{rajput2023tiger,zheng2025mtgrec}.
Semantic-identifier models are generative too, but over discrete item codes.
In contrast, DARTS samples continuous next-item representations, so the same history can produce different rankings.

\paragraph{Diffusion models for recommendation.}
Diffusion recommenders aim at generating conditional recommendation distribution. They fall into two families.
DiffRec~\citep{wang2023diffrec} and DDRM~\citep{zhao2024ddrm} diffuse a user's interaction vector, while the second family, which includes DARTS, diffuses a representation of the target item conditioned on the history.
In this second family, DiffuRec~\citep{li2023diffurec} denoises the target embedding with a transformer, DreamRec~\citep{yang2023dreamrec} adds classifier-free guidance, and ADRec~\citep{chen2025adrec} warm-starts its denoiser from collaborative embeddings.
\citet{zolghadr2024diffurecsys} study the effect of offset noise and cross-attention.
Other methods change the training signal or the corruption process.
PreferDiff~\citep{liu2025preferdiff} derives a ranking loss compatible with diffusion, and PreferGrow~\citep{hu2025prefergrow} diffuses discrete preference ratios.
Most target-item methods learn the item representation together with the denoiser, and many add item-level losses during denoising.
DARTS freezes a pretrained latent space, trains the denoiser by latent regression only, and applies ranking supervision only to the final decoder.

\paragraph{Latent and continuous generative modelling.}
Latent diffusion~\citep{rombach2022ldm} separates representation learning from generation by training and freezing an autoencoder first, then fitting the diffusion model in its latent space.
In language, Diffusion-LM~\citep{li2022diffusionlm} diffuses in embedding space but repeatedly projects back onto tokens.
ELF~\citep{hu2026elf} instead stays continuous until a single final discretisation, and shows that this improves both generation quality and efficiency.
We bring both ideas, a fixed latent space and one final discretisation, to recommendation, and test whether they keep sampling stochastic when tokens are replaced by catalogue items.

%% file: content/method.tex
\section{Method}
\label{sec:method}

\subsection{Overview}
\label{ssec:overview}

DARTS is composed by three architectural components: an encoder, a decoder, and
a denoising diffusion model. The training of these components is realized in three separated stages that are key for a diffusion model for recommendation :
the representation learning of the catalogue item in a latent space,  the latent generation, and  the item ranking.
Each stage updates a different set of parameters of the model.

\begin{enumerate}
  \item \textbf{Sequence autoencoder pretraining.} The encoder and decoder are trained with masked item modelling~\citep{devlin2019bert,sun2019bert4rec}, which defines a latent space tied to the item catalogue.
  \item \textbf{Latent diffusion.} With the encoder and decoder frozen, a conditional denoiser learns to recover the latent of the next item from its noisy version, using the user history as context.
  \item \textbf{Decoder finetuning.} With the encoder and denoiser frozen, the decoder is finetuned on sampled latents to rank the true next item first, which corrects the shift between the latents it was pretrained on and those generated at inference.
\end{enumerate}

Figure~\ref{fig:architecture} gives an overview, and Appendix~\ref{app:training-stages} summarises the parameters trained at each stage.

\input{figures/architecture.tex}

\subsection{Background: conditional denoising diffusion}
\label{ssec:preliminaries}
To model several plausible continuations of an interaction history, we use conditional diffusion \cite{ho2020ddpm} to learn a distribution over next-item latents. and provide below a brief background.

\paragraph{Forward process.}
Given a clean vector $\vz^{0}$ sampled from a target distribution, diffusion defines a fixed Markov process $q$ that gradually adds Gaussian noise over $T$ steps~\citep{ho2020ddpm}.
Let $\beta_1,\ldots,\beta_T$ be the variance schedule, $\alpha_t=1-\beta_t$, and $\bar{\alpha}_t=\prod_{r=1}^{t}\alpha_r$.
A noisy vector at any step $t$ can be sampled directly as
\begin{equation}
  \label{eq:forward}
  \vz^{t}
  =
  \sqrt{\bar{\alpha}_t}\,\vz^{0}
  +\sqrt{1-\bar{\alpha}_t}\,\bm{\epsilon},
  \qquad
  \bm{\epsilon}\sim\mathcal{N}(\vzero,\mI).
\end{equation}
where increasing $t$ removes information from $\vz^{0}$.
The schedule is chosen so that $\vz^{T}$ is approximately distributed as $\mathcal{N}(\vzero,\mI)$.

\paragraph{Reverse process.}
Starting from Gaussian noise $\vz^{T}$, generation iteratively produces $\vz^{T-1},\ldots,\vz^{0}$.
Under the $\vz^{0}$ parameterisation, the denoiser directly estimates the clean vector as $\hat{\vz}^{0}=g_\theta(\vz^{t},t,\vc)$ from the current noisy vector, timestep, and condition $\vc$.
Substituting this estimate into the Gaussian posterior of the forward process gives the ancestral update
\begin{equation}
  \label{eq:reverse}
  \vz^{t-1}
  \quad=\quad
  \frac{\sqrt{\bar{\alpha}_{t-1}}\, \beta_t}{1 - \bar{\alpha}_t}\, \hat{\vz}^{0}
  + \frac{\sqrt{\alpha_t}\,(1 - \bar{\alpha}_{t-1})}{1 - \bar{\alpha}_t}\, \vz^{t}
  + \sqrt{\tilde{\beta}_t}\, \bm{\epsilon},
  \qquad
  \tilde{\beta}_t = \frac{1 - \bar{\alpha}_{t-1}}{1 - \bar{\alpha}_t}\, \beta_t ,
\end{equation}
where $\bm{\epsilon} \sim \mathcal{N}(\vzero,\mI)$ for $t>1$ and $\bm{\epsilon}=\vzero$ for the final step.
During training, a timestep is sampled according to a chosen schedule and the denoiser learns to recover $\vz^{0}$ from $\vz^{t}$ and $\vc$.

In  our context of diffusion recommendation,
existing sequential diffusion recommenders \cite{li2023diffurec,wang2023diffrec} often use a learned item embedding as $\vz^{0}$ and update the embedding table with the denoiser. We have
a different strategy and we will clarify later the target distribution from which $\vz^0$ is sampled and the condition vector.


\subsection{Item representations and model backbone}
\label{ssec:setup}


Let $\mathcal{I}$ be the discrete item catalogue of size $N$ and let
$s=(i_1,\ldots,i_L)\in\mathcal{I}^L$ be a user's chronologically ordered interaction sequence of length $L$.
For a sake of clarity, we will assume that $L$ is fixed for all sequences.
Given a binary mask $m\in\{0,1\}^L$, its corrupted counterpart is
$\tilde{s}=(\tilde{i}_1,\ldots,\tilde{i}_L)\in
(\mathcal{I}\cup\{\texttt{[MASK]}\})^L$, where
\[
  \tilde{i}_t =
  \begin{cases}
    \texttt{[MASK]}, & m_t=1,\\
    i_t,             & m_t=0.
  \end{cases}
\]
The recommendation task is to predict $i_L$ from the clean prefix $i_{1:L-1}$. Note that if the sequence length is
smaller than $L$, then the sequence is left-padded to $L$.

We distinguish the discrete catalogue $\mathcal{I}$ from two continuous vector spaces:
the item representation space $\mathcal{V}=\mathbb{R}^{d_v}$, in which the catalogue item representations $\ve_i$ live, and the latent space $\mathcal{Z}=\mathbb{R}^{d_z}$, in which the encoder representation and diffusion process resides.

\paragraph{Hybrid item representations.}
Each item $i\in\mathcal{I}$ has a representation vector $\ve_i\in\mathcal{V}$ that combines semantic and collaborative information:
\begin{equation}
  \label{eq:item-embedding}
  \ve_i
  =
  \omega\,\ve_i^{\mathrm{sem}}
  +(1-\omega)\,\ve_i^{\mathrm{id}}.
\end{equation}
The semantic vector $\ve_i^{\mathrm{sem}}$ is a frozen \href{https://huggingface.co/sentence-transformers/sentence-t5-base}{\texttt{sentence-t5-base}} encoding of the item's title and description~\citep{ni2022sentencet5}. The identifier vector $\ve_i^{\mathrm{id}} \in \mathbb{R}^{d_v}$ is learned during Stage~1, and the mixing weight $\omega$ is a hyperparameter. We reuse the mixed item matrix as the decoder output projection, tying the input representations to the catalogue scores~\citep{press2017tying}. Mixing the semantic and the collaborative information improved the performance, see ablation in \ref{app:t5-ablation}.

\paragraph{Encoder and decoder.}
The bidirectional transformer encoder $E_\phi$ maps a sequence of $L$ item tokens to a sequence of $L$ contextual states in $\mathcal{Z}$:
\[
  E_\phi(s)=\vz_{1:L},
  \qquad
  \vz_j\in\mathcal{Z}.
\]
Depending on the training stage, the input of the encoder may be a corrupted one $\tilde s$ (during autoencoder training) or a clean sequence $s$ (
when training the denoising model).
The decoder $D_\psi$ is a causal transformer that takes as input a latent sequence and outputs a sequence of items in the item representation space
$\mathcal{V}$.



Hence, diffusion operates in the latent space $\mathcal{Z}$ , while the decoder outputs a sequence of item representation in $\mathcal{V}$.
Figure~\ref{fig:architecture} shows how the components fit together. Appendix~\ref{app:blocks} gives the block diagrams.

\subsection{Stage 1: sequence autoencoder pretraining}
\label{ssec:stage1}

\paragraph{Objective.}
We train $(E_\phi, D_\psi)$ with masked item modeling.
Given a training sequence $s=(i_1,\ldots,i_L)$, we sample a Bernoulli mask $m \in \{0,1\}^L$ with rate $p$ used to form the corrupted sequence $\tilde{s}=(\tilde{i}_1,\ldots,\tilde{i}_L)$.

We denote as $ \vh_{1:L} = [\vh_1, \cdots, \vh_L] $, with $\vh_j \in \mathcal{V}$,  the outputs of the autoencoder $ D_\psi\!\left(E_\phi(\tilde{s})\right)$ and we train the  couple to recover the original items at the masked positions. Given
\[
  p_{\phi,\psi}\!\left(i\mid\tilde{s},j\right)
  =
  \frac{\exp\!\left(\vh_j^{\top}\ve_i\right)}
       {\sum_{i'\in\mathcal{I}}\exp\!\left(\vh_j^{\top}\ve_{i'}\right)}.
\]
$p_{\phi,\psi}$ being the catalogue-wide softmax distribution induced at position $j$ by the encoder--decoder pair and the item embeddings,
we consider the following loss for the autoencoder training
\begin{equation}
  \label{eq:mlm}
  \mathcal{L}_{\text{MLM}}(\phi, \psi)
  \quad=\quad
  -\,\E_{s,m}\left[
  \frac{1}{|\mathcal{M}|}
  \sum_{j\in\mathcal{M}}
  \log p_{\phi,\psi}\!\left(i_j\mid\tilde{s},j\right)
  \right].
\end{equation}

\paragraph{Choice of masking ratio.}
We use a masking ratio of $0.4$ and always mask the final item to match the next-item prediction objective of sequential recommendation.

After Stage\~1, the encoder and item representations remain frozen during Stages\~2 and\~3. The decoder is kept frozen during Stage\~2 and fine-tuned during Stage\~3.

\subsection{Stage 2: latent diffusion}
\label{ssec:stage2}

In this stage, we train the diffusion denoiser to model the distribution of the next-item latent conditioned on the observed interaction history, while keeping the pretrained latent space fixed.

\paragraph{Condition and target.}
Diffusion training uses two encoder passes with different inputs. First, we mask the final item:
\[
  s^{\mathrm{mask}}
  =
  (i_1,\ldots,i_{L-1},\texttt{[MASK]}).
\]
and encode  this sequence to obtain  the conditioning prefix of the diffusion denoiser
\begin{equation}
  \label{eq:diffusion-condition}
  \vc
  =
  \left[E_\phi(s^{\mathrm{mask}})\right]_{1:L-1}.
\end{equation}

The target distribution must relate to the distribution of final item $i_L$. To define the clean diffusion target $\vz0$, we pass the uncorrupted sequence $s$ through $E_\phi$ and extract the representation at the final slot:
\begin{equation}
  \label{eq:diffusion-target}
  \vz^0 = \operatorname{sg}\!\left(\left[E_\phi(s)\right]_L\right),
\end{equation}
where $\operatorname{sg}$ denotes the stop-gradient operator. Consequently, the condition $\vc$ uses strictly the information available at inference time, while the target $\vz^0$ reflects the contextualized representation of the target item given the full sequence.

This key formulation differs from existing target-item diffusion approaches~\citep{li2023diffurec,yang2023dreamrec,chen2025adrec}, which diffuse isolated, static item embeddings $e(i_L)$. By contrast, our target $\vz^0$ is generated by the sequence encoder $E_\phi$, capturing rich context-dependent item semantics. Furthermore, because $E_\phi$ is frozen, both the conditioning space and the target space remain fixed throughout diffusion training, preventing representation drift.

\paragraph{Denoiser's Architecture.}
The diffusion denoiser $g_\theta$ is a stack of transformer blocks. At each block, the noisy target latent $\vz^t$ acts as the cross-attention query and the prefix states $\vc$ provide keys and values. The timestep modulates the block through adaptive layer normalisation~\citep{peebles2023dit}. Padding positions in the prefix are excluded from cross-attention.

\paragraph{Objective.}
For a sampled timestep $t$ and noise $\bm{\epsilon}\sim\mathcal{N}(\vzero,\mI)$, we construct
\[
  \vz^{t} = \sqrt{\bar{\alpha}_t}\,\vz^{0} +\sqrt{1-\bar{\alpha}_t}\,\bm{\epsilon}.
\]
 The denoiser is trained with latent root mean squared error and predicts the clean target directly.
\begin{equation}
  \label{eq:diffusion}
  \mathcal{L}_{\mathrm{diff}}(\theta)
  =
  \sqrt{
    \E_{s,t,\bm{\epsilon}}\left[
      \frac{1}{d_z}
      \left\|
        g_\theta(\vz^{t},t,\vc)-\vz^{0}
      \right\|_2^2
    \right]
  }.
\end{equation}
We want to emphasize that contrarily existting approaches, no catalogue-related loss is applied in this stage.

\subsection{Stage 3: Decoder Fine-Tuning}
\label{ssec:stage3}

In Stage 1, the encoder and decoder are trained jointly to reconstruct the full item sequence. At inference, however, the decoder receives the conditioning sequence $\vc$ of length $L-1$ and the generated target latent $\hat{\vz}_0$ at the final slot. This creates a distribution shift: slot $L$ contains a masked context representation during pretraining but a latent sample from reverse diffusion at inference. Stage 3 closes this gap by finetuning the decoder to rank the target item above candidates using its inference inputs. The encoder $E_\phi$, denoiser $g_\theta$, and item embeddings remain frozen.

For a training sequence $s$, we extract the conditioning  $\vc$ via Equation~\ref{eq:diffusion-condition}, sample $\vz_T \sim \mathcal{N}(\vzero, \mI)$, and execute the full reverse diffusion process to sample a target latent $\hat{\vz}_0 \sim p_\theta(\vz_0 \mid \vc)$. We concatenate $\vc$ and $\hat{\vz}_0$ to form the $L$-length input sequence for the decoder $D_\psi$. The output of interest is the logit vector $\vl \in \mathbb{R}^N$ over the item catalogue at the final position.

Let $y = i_L$ be the ground-truth target item, and let $\mathcal{H}_n(\vl, y)$ denote the set of $n$ highest-scoring negative items (i.e., self-mined hard negatives). We fine-tune $D_\psi$ using a temperature-scaled Bayesian Personalized Ranking (BPR) loss~\citep{rendle2009bpr}:
\begin{equation}
  \label{eq:decoder}
  \mathcal{L}_{\mathrm{dec}}(\psi)
  =
  -\frac{1}{n}
  \sum_{j \in \mathcal{H}_n(\vl, y)}
  \log \sigma\!\left( \frac{l_y - l_j}{\tau} \right),
\end{equation}
where $\tau > 0$ is a temperature hyperparameter and $\sigma(\cdot)$ is the sigmoid function. Optimizing $\mathcal{L}_{\mathrm{dec}}$ adapts the decoder to map continuous sampled latents directly to discrete catalogue items, resolving the inference distribution shift without changing the stochastic latent distribution learned in Stage 2.
Inference follows the same sampling and decoding path (Appendix~\ref{app:inference}).

%% file: figures/architecture.tex
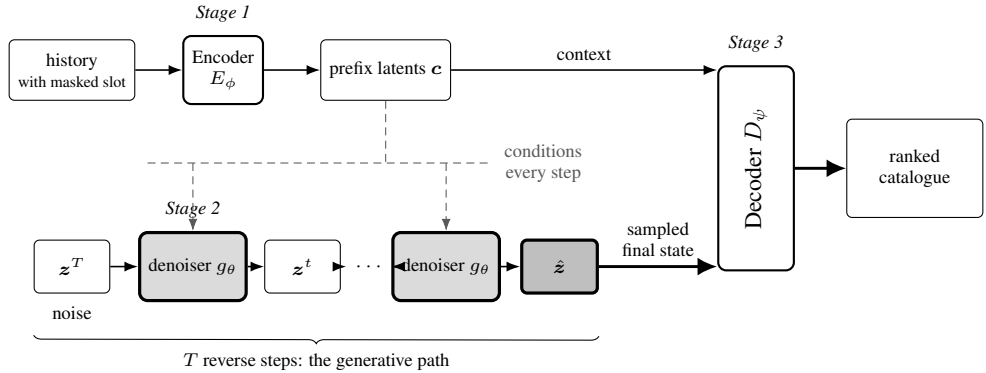
\begin{figure}[!htbp]
  \centering
  \begin{tikzpicture}[
      font=\small,
      tok/.style={draw, rounded corners=2pt, minimum width=15mm, minimum height=8mm,
                  align=center, font=\scriptsize},
      state/.style={draw, rounded corners=2pt, minimum width=10mm, minimum height=7mm,
                    inner sep=1pt, font=\scriptsize},
      gen/.style={draw, very thick, rounded corners=3pt, fill=black!14,
                  minimum width=14mm, minimum height=9mm, align=center, font=\scriptsize},
      genout/.style={state, very thick, fill=black!25},
      mod/.style={draw, thick, rounded corners=3pt, minimum width=10mm,
                  minimum height=9mm, align=center, font=\scriptsize},
      tall/.style={draw, thick, rounded corners=3pt, minimum width=10mm,
                   minimum height=27mm, align=center},
      ar/.style={-{Latex[length=1.8mm]}, semithick},
      fat/.style={-{Latex[length=2.6mm, width=2.4mm]}, line width=1.1pt},
      cnd/.style={-{Latex[length=1.6mm]}, densely dashed, gray!70!black},
      lbl/.style={font=\scriptsize, align=center},
      tag/.style={font=\scriptsize\itshape, align=center},
    ]
    \def\ytop{1.65} \def\ybus{0.42} \def\ygen{-0.95}

    \node[tok] (hist) at (0.35,\ytop) {history\\{\tiny with masked slot}};
    \node[mod] (enc)  at (2.35,\ytop) {Encoder\\$E_\phi$};
    \node[tok, minimum width=17mm] (ctx) at (4.50,\ytop) {prefix latents $\vc$};
    \draw[ar] (hist) -- (enc);
    \draw[ar] (enc) -- (ctx);
    \node[tag, above=0.5mm of enc] {Stage 1};

    \draw[gray!70!black, densely dashed] (ctx.south) -- (4.50,\ybus);
    \draw[gray!70!black, densely dashed] (1.35,\ybus) -- (5.85,\ybus);
    \node[lbl, text=gray!70!black, anchor=west] at (5.95,\ybus)
      {conditions\\every step};

    \node[state] (zT)  at (0.35,\ygen) {$\vz^{T}$};
    \node[gen]   (g1)  at (1.95,\ygen) {denoiser $g_\theta$};
    \node[state] (zt)  at (3.40,\ygen) {$\vz^{t}$};
    \node[lbl]   (dts) at (4.30,\ygen) {$\cdots$};
    \node[gen]   (g2)  at (5.30,\ygen) {denoiser $g_\theta$};
    \node[genout](zh)  at (6.80,\ygen) {$\hat{\vz}$};

    \node[lbl, below=0.5mm of zT] {noise};
    \node[tag, above=0.5mm of g1] {Stage 2};

    \draw[ar] (zT) -- (g1);
    \draw[ar] (g1) -- (zt);
    \draw[ar] (zt) -- (dts);
    \draw[ar] (dts) -- (g2);
    \draw[ar] (g2) -- (zh);

    \foreach \n in {g1,g2} { \draw[cnd] (\n.north |- 0,\ybus) -- (\n.north); }

    \draw[decorate, decoration={brace, mirror, amplitude=3pt}]
      ($(zT.south west)+(0,-0.55)$) -- ($(zh.south east)+(0,-0.55)$)
      node[midway, below=3pt, lbl] {$T$ reverse steps: the generative path};

    \node[tall] (dec) at (9.40,0.35) {\rotatebox{90}{Decoder $D_\psi$}};
    \node[tag, above=0.5mm of dec] {Stage 3};
    \node[tok, minimum width=18mm, minimum height=13mm] (sc) at (11.50,0.35)
      {ranked\\catalogue};

    \draw[ar]  (ctx.east) -- (dec.west |- ctx.east)
      node[midway, above, lbl] {context};
    \draw[fat] (zh.east) -- (dec.west |- zh.east)
      node[midway, above, lbl] {sampled\\final state};
    \draw[fat] (dec) -- (sc);
  \end{tikzpicture}
  \caption{Model overview. The history is encoded once with its final slot masked. The prefix latents $\vc$ condition every reverse step. Starting from noise, the denoiser runs $T$ times to produce the sampled final state $\hat{\vz}$. The decoder combines $\hat{\vz}$ with the fixed prefix to rank the catalogue.}
  \label{fig:architecture}
\end{figure}

%% file: content/experiments.tex
\section{Experiments}
\label{sec:experiments}

In this section, we empirically evaluate DARTS against competitive baselines. We further introduce diagnostic measurements to analyze the stochastic properties of the diffusion process and quantify the diversity of its predictions. We also conduct controlled ablations to isolate how common design choices in prior work induce sampling collapse.

\subsection{Datasets}
\label{ssec:datasets}

We evaluate DARTS on three categories of the Amazon~2023 review corpus~\citep{hou2024amazon2023}: \textsc{Musical Instruments} (Instrument), \textsc{Industrial and Scientific} (Scientific), and \textsc{Video Games} (Game).
Following prior work~\citep{rajput2023tiger,zhou2020s3rec,zheng2025mtgrec}, we retain users and items with at least five interactions, order each history chronologically, and keep its $20$ most recent items.
Appendix Table~\ref{tab:dataset-stats} reports the resulting statistics.

\subsection{Overall Performance - Accuracy}
\label{ssec:overall-performance}

Table~\ref{tab:main-results} compares DARTS against three families of baselines: sequential recommenders~\citep{hidasi2016gru4rec,sun2019bert4rec,kang2018sasrec,zhou2022fmlprec,zhai2024hstu,zhang2019fdsa,zhou2020s3rec}, generative retrieval with semantic identifiers~\citep{rajput2023tiger,wang2024letter,hou2025actionpiece}, and models we train ourselves with the same hybrid item representations as DARTS (\emph{+T5}): SASRec and the diffusion recommenders DreamRec~\citep{yang2023dreamrec}, DiffuRec~\citep{li2023diffurec}, and ADRec~\citep{chen2025adrec}.
Results for the first two families are taken from~\citet{zheng2025mtgrec}, except ActionPiece from~\citet{zhong2025pctx}; all follow the protocol above.
We report Recall@$K$ and NDCG@$K$ for $K \in \{5, 10\}$ under leave-one-out evaluation; Appendix~\ref{app:implementation} details baselines, implementation, and hyperparameters.

\input{tables/main-results.tex}

In terms of these accuracy measures, DARTS performs best on all twelve combinations of datasets and metrics, improving over the strongest baseline in each column by between $6.8\%$ and $12.0\%$.

\paragraph{Disentangling item representations from the generative formulation.}
SASRec\,+\,T5 shares our item representations and codebase, so this comparison isolates the generative process. Our NDCG@10 gains are $11.9\%$ on \textsc{Instrument}, $16.4\%$ on \textsc{Scientific}, and $19.5\%$ on \textsc{Game} w.r.t. SASRec\,+\,T5.
The diffusion baselines share the same representations yet vary widely, so our gains come from the staged design rather than from item features.
Appendix~\ref{app:t5-ablation} isolates the representation type. Semantic features lift our NDCG@10 by $14.4\%$, but our identifier-only variant still leads every identifier-only competitor by $18.6\%$.
DiffuRec, the closest baseline since it also diffuses target representations, learns its latent space jointly with the denoiser and applies intermediate catalogue cross-entropy, two design choices we further evaluate next.

\subsection{Assessing the diversity-accuracy coompromise}
\label{ssec:generativeness}

Single-draw ranking metrics cannot determine whether diffusion produces meaningful stochastic variation.
For each test context $\vc$, we therefore draw $K=10$ independent latents
$\hat \vz^0$ from $p_\theta(\cdot\mid\vc)$, decode each into a ranked list, and compare the draws.
\paragraph{Diagnostics.} We introduce three metrics that help
us evaluate the generative capability of generative recommender
models in terms of diversity of the predictions.

The first one is a mean pairwise Jaccard dissimilarity measure. Assuming that  $\mathcal{S}^{(k)}_{\vc}$ denotes the top-$10$ set from draw $k$,
we have the measure
\[
J = \E_{\vc}\!\left[\binom{K}{2}^{-1}
\sum_{k<l}\left(1-\frac{|\mathcal{S}^{(k)}_{\vc}\cap\mathcal{S}^{(l)}_{\vc}|}
{|\mathcal{S}^{(k)}_{\vc}\cup\mathcal{S}^{(l)}_{\vc}|}\right)\right].
\]
$J$ is zero for identical slates and grows as resampling yields distinct recommendations. Because $J$ is computed after decoding, it reflects the combined behavior of the sampler and the decoder. The next two metrics operate in latent space and isolate the stochasticity of the denoiser.

Repeated samples may fail to be generative in two ways: they may exhibit negligible variation or explore only a few latent directions. We measure these complementary forms of collapse using variance decomposition and effective dimensionality. For $K$ draws $\{\hat{\vz}_{\vc}^{(k)}\}_{k=1}^K$, let $\bar{\vz}_{\vc}=K^{-1}\sum_k\hat{\vz}_{\vc}^{(k)}$ and $\bar{\vz}=\E_{\vc}[\bar{\vz}_{\vc}]$. Define $V_{\mathrm{w}}=\E_{\vc}[K^{-1}\sum_k\|\hat{\vz}_{\vc}^{(k)}-\bar{\vz}_{\vc}\|_2^2]$ and $V_{\mathrm{b}}=\E_{\vc}[\|\bar{\vz}_{\vc}-\bar{\vz}\|_2^2]$. By the within--between variance decomposition, the \emph{generative variance ratio}
\begin{equation}
  \label{eq:variance-ratio}
  R=\frac{V_{\mathrm{w}}}{V_{\mathrm{w}}+V_{\mathrm{b}}}
\end{equation}
is the fraction of latent variance due to resampling rather than differences between histories. $R=0$ indicates a deterministic point predictor.

As $R$ quantifies the magnitude of sampling variation, but not how many directions it spans, we therefore define the \emph{participation ratio} of the within-context sample covariance matrix $\mC_{\vc}$:
\begin{equation}
  \label{eq:participation-ratio}
  \mathrm{PR} = \E_{\vc} \left[ \frac{\left( \operatorname{tr} \mC_{\vc} \right)^2}{\operatorname{tr} \left( \mC_{\vc}^2 \right)} \right],
\end{equation}
where $\mC_{\vc} = \frac{1}{K-1} \sum_{k=1}^K (\hat{\vz}_{\vc}^{(k)} - \bar{\vz}_{\vc})(\hat{\vz}_{\vc}^{(k)} - \bar{\vz}_{\vc})^\top$. The participation ratio estimates the effective dimensionality of the sampled subspace: approaching $0$ for a completely collapsed point, $1$ when variance is restricted to a single axis, and up to $\min(K-1, d)$ for isotropic exploration across available dimensions.

\paragraph{Results.}Neither $R$ nor $\mathrm{PR}$ measures recommendation quality, so we interpret them alongside NDCG and the decoded-set diagnostic $J$.
Table~\ref{tab:generative-metrics} reports the three diagnostics on \textsc{Scientific}.
All three baselines produce nearly fixed slates, but their latent diagnostics reveal different failure modes.
ADRec collapses in both latent and item space.
DiffuRec explores several directions ($\mathrm{PR}=5.0$), but at negligible scale relative to variation across histories ($R=4.8\times10^{-5}$), so $98\%$ of its top-$10$ is preserved between draws.
DreamRec has a larger variance ratio ($R=0.08$) but an almost zero participation ratio ($0.005$), indicating little latent exploration. Its low NDCG further shows why these diagnostics cannot be read as quality measures.

\input{tables/generative-metrics.tex}

DARTS combines variation ($R=0.30$) with multi-directional exploration ($\mathrm{PR}=4.7$), and this diversity survives decoding: $J=0.30$, i.e.\ a pairwise Jaccard similarity of $0.70$ between top-$10$ sets. It thus achieves strong diversity while outperforming all baselines in Table~\ref{tab:main-results}.

\subsection{Ablations}
\label{ssec:ablations}

\paragraph{Catalogue cross-entropy supervision should not be used inside the diffusion process.}
We retrain Stage~2 from one Stage-1 checkpoint with $(1-\alpha)\mathcal{L}_{\text{diff}}+\alpha\mathcal{L}_{\text{CE}}$,
where   $\mathcal{L}_{\text{CE}}$ is defined in Equation ~\ref{eq:ce-ablation}
and compute a loss at the decoder level. The setting $\alpha=0$ is our latent-only objective, while $\alpha=1$ uses cross-entropy alone.
Results and detailed analyses are in Table \ref{tab:ce-ablation}
and Appendix~\ref{app:ce-ablation}.
We note that even $\alpha = 0.1$ reduces $R$ by more than half and lowers $PR$ from 4.7 to 1.5. Both metrics continue to decrease as $\alpha$ increases, reaching 0.005 and 0.65, respectively, at $\alpha = 1$.
In addition, the use of  cross-entropy loss  does not improve ranking either: NDCG@5 is highest at $\alpha=0$ and decreases monotonically to $0.0089$ at $\alpha=1$, a drop of $57\%$.
Discrete supervision therefore suppresses generative properties without improving ranking accuracy. This finding is  consistent with the continuous-space argument of \citet{hu2026elf}.

\paragraph{Freezing the item representations.}
We next test whether the item representations should remain fixed during Stage~2 (Details in Appendix~\ref{app:design-ablation} and Table \ref{tab:design-ablation}).
Under latent regression loss, updating them jointly with the denoiser reduces $R$ from $0.300$ to $0.024$, lowers $\mathrm{PR}$ from $4.70$ to $1.30$, and decreases NDCG@5 by $42\%$.
Thus,  keeping  fix the item representations and excluding catalogue cross-entropy  loss are necessary to obtain the strong sampling behaviour and ranking accuracy.

\paragraph{Generative behaviour converts into accuracy.}
Table~\ref{tab:generative-metrics} shows that DARTS is generative and Table~\ref{tab:main-results} that a single draw is accurate; we now show that aggregating multiple draws turns this diversity into further accuracy gains.
DARTS allows us to define the posterior-predictive mixture $p(i | \vc)=\E_{\vz\sim p_\theta}[p_\psi(i | \vz,\vc)]$,
where $p_\psi(i | \vz,\vc)$ is the softmax over item scores. We estimate $p(i | \vc)$ by averaging $p_\psi$ over $M$ draws  and rank items by this average at a fixed top-$k$.
Note that a model that returns the same item distribution on every pass cannot benefit from repeated evaluation because identical distributions aggregate to themselves.
\input{figures/power-mean-ndcg10.tex}
Evaluating this aggregation on \textsc{Scientific}
(see Figure~\ref{fig:power-mean-ndcg10} and the full results in Appendix \ref{app:inference-time} Figure~\ref{fig:power-mean}), DARTS improves on all four metrics from $M=1$ to $M=4$, by $3.2\%$ on both NDCG@5 and NDCG@10, while DiffuRec moves by at most $0.6\%$ and is effectively flat beyond $M=2$.
Thus, additional draws improve our ranking at fixed slate size without retraining, whereas DiffuRec's nearly coincident draws provide no benefit.

Each draw repeats the reverse chain, so this gain is not free: throughput in predictions per second falls by $3.3\times$ between $M=1$ and $M=4$.
Appendix~\ref{app:inference-time} reports end-to-end inference cost and shows that the $M=1$ model is nonetheless $3.6\times$ faster than DiffuRec with better ranking accuracy, because our denoiser conditions on a single latent token instead of re-encoding the item sequence at every step.
Appendix~\ref{app:union-recall} contrasts this with pooling the top-$n$ lists, which does not improve accuracy once the larger slate is accounted for.
Finally, randomness and repeated inference alone do not explain these gains.
Perturbing SASRec's catalogue distribution with Gumbel noise sharply reduces ranking accuracy, and aggregating up to $M=15$ draws only moves the result back toward, and never above, the deterministic ranking (Appendix~\ref{app:stochastic-sasrec} and Figure~\ref{fig:stochastic-sasrec}).

%% file: tables/main-results.tex
\begin{table}[!htbp]
    \caption{Recommendation performance on three Amazon Reviews 2023 categories. R@$K$/N@$K$ are Recall/NDCG@$K$; \textbf{best} and \underline{second best} are marked, and \emph{Improv.}\ is our gain over the best baseline per column.}
    \label{tab:main-results}
    \centering
    \small
    \setlength{\tabcolsep}{3pt}
    \resizebox{\textwidth}{!}{%
    \begin{tabular}{lcccccccccccc}
        \toprule
        & \multicolumn{4}{c}{Instrument} & \multicolumn{4}{c}{Scientific} & \multicolumn{4}{c}{Game} \\
        \cmidrule(lr){2-5} \cmidrule(lr){6-9} \cmidrule(lr){10-13}
        Method & R@5 & R@10 & N@5 & N@10 & R@5 & R@10 & N@5 & N@10 & R@5 & R@10 & N@5 & N@10 \\
        \midrule
        \multicolumn{13}{l}{\emph{Sequential recommenders}} \\
        GRU4Rec  & 0.0324 & 0.0501 & 0.0209 & 0.0266 & 0.0202 & 0.0338 & 0.0129 & 0.0173 & 0.0499 & 0.0799 & 0.0320 & 0.0416 \\
        BERT4Rec & 0.0307 & 0.0485 & 0.0195 & 0.0252 & 0.0186 & 0.0296 & 0.0119 & 0.0155 & 0.0460 & 0.0735 & 0.0298 & 0.0386 \\
        SASRec   & 0.0333 & 0.0523 & 0.0213 & 0.0274 & 0.0259 & 0.0412 & 0.0150 & 0.0199 & 0.0535 & 0.0847 & 0.0331 & 0.0438 \\
        FMLP-Rec & 0.0339 & 0.0536 & 0.0218 & 0.0282 & 0.0269 & 0.0422 & 0.0155 & 0.0204 & 0.0528 & 0.0857 & 0.0338 & 0.0444 \\
        HSTU     & 0.0343 & 0.0577 & 0.0191 & 0.0271 & 0.0271 & 0.0429 & 0.0147 & 0.0198 & 0.0578 & 0.0903 & 0.0334 & 0.0442 \\
        FDSA     & 0.0347 & 0.0545 & 0.0230 & 0.0293 & 0.0262 & 0.0421 & 0.0169 & 0.0213 & 0.0544 & 0.0852 & 0.0361 & 0.0448 \\
        S\textsuperscript{3}-Rec & 0.0317 & 0.0496 & 0.0199 & 0.0257 & 0.0263 & 0.0418 & 0.0171 & 0.0219 & 0.0485 & 0.0769 & 0.0315 & 0.0406 \\
        \midrule
        \multicolumn{13}{l}{\emph{Generative retrieval with semantic identifiers}} \\
        TIGER       & 0.0370 & 0.0564 & 0.0244 & 0.0306 & 0.0264 & 0.0422 & 0.0175 & 0.0226 & 0.0559 & 0.0868 & 0.0366 & 0.0467 \\
        LETTER      & 0.0372 & 0.0580 & 0.0246 & 0.0313 & 0.0279 & 0.0435 & 0.0182 & 0.0232 & 0.0563 & 0.0877 & 0.0372 & 0.0473 \\
        ActionPiece & 0.0383 & \underline{0.0615} & 0.0243 & \underline{0.0318} & 0.0284 & 0.0452 & 0.0182 & 0.0236 & \underline{0.0591} & 0.0927 & \underline{0.0382} & \underline{0.0490} \\
        \midrule
        \multicolumn{13}{l}{} \\
        SASRec + T5   & 0.0376 & 0.0595 & 0.0243 & 0.0311 & 0.0277 & 0.0452 & 0.0177 & 0.0232 & 0.0534 & 0.0853 & 0.0344 & 0.0447 \\
        DreamRec + T5 & 0.0146 & 0.0269 & 0.0088 & 0.0124 & 0.0112 & 0.0196 & 0.0071 & 0.0098 & 0.0146 & 0.0249 & 0.0086 & 0.0122 \\
        DiffuRec + T5 & \underline{0.0387} & 0.0596 & \underline{0.0248} & 0.0315 & \underline{0.0302} & \underline{0.0468} & \underline{0.0190} & \underline{0.0243} & \underline{0.0591} & \underline{0.0944} & 0.0368 & 0.0482 \\
        ADRec + T5    & 0.0314 & 0.0498 & 0.0201 & 0.0260 & 0.0236 & 0.0374 & 0.0150 & 0.0195 & 0.0549 & 0.0869 & 0.0357 & 0.0459 \\
        \midrule
        \textbf{DARTS} & \textbf{0.0423} & \textbf{0.0663} & \textbf{0.0271} & \textbf{0.0348} & \textbf{0.0329} & \textbf{0.0524} & \textbf{0.0207} & \textbf{0.0270} & \textbf{0.0653} & \textbf{0.1045} & \textbf{0.0408} & \textbf{0.0534} \\
        \midrule
        \emph{Improv.} & +9.3\% & +7.8\% & +9.3\% & +9.4\% & +8.9\% & +12.0\% & +8.9\% & +11.1\% & +10.5\% & +10.7\% & +6.8\% & +9.0\% \\
        \bottomrule
    \end{tabular}%
    }
\end{table}

%% file: tables/generative-metrics.tex
\begin{table}[!htbp]
    \caption{Sampling diagnostics on \textsc{Scientific} from $K = 10$ draws per context: Jaccard dissimilarity $J$, generative variance ratio $R$, and participation ratio $\mathrm{PR}$. NDCG@5 (single draw) is repeated to show the diagnostics are not a proxy for ranking quality.}
    \label{tab:generative-metrics}
    \centering
    \small
    \begin{tabular}{lcccc}
        \toprule
        Model & $J$ $\uparrow$ & $R$ $\uparrow$ & $\mathrm{PR}$ $\uparrow$ & NDCG@5 \\
        \midrule
        DreamRec\,+\,T5 & $0.020$ & $8.0 \times 10^{-2}$ & $0.005$ & $0.0071$ \\
        DiffuRec\,+\,T5 & $0.019$ & $4.8 \times 10^{-5}$ & $5.0$ & $0.0190$ \\
        ADRec\,+\,T5    & $3.0 \times 10^{-4}$ & $9.0 \times 10^{-9}$ & $2.6 \times 10^{-5}$ & $0.0150$ \\
        \midrule
        \textbf{DARTS}  & $\mathbf{0.30}$ & $\mathbf{0.30}$ & $4.7$ & $\mathbf{0.0207}$ \\
        \bottomrule
    \end{tabular}
\end{table}

%% file: figures/power-mean-ndcg10.tex
\begin{wrapfigure}{!htbp}{0.47\textwidth}
  \centering
  \vspace{-\baselineskip}
  \includegraphics[width=\linewidth]{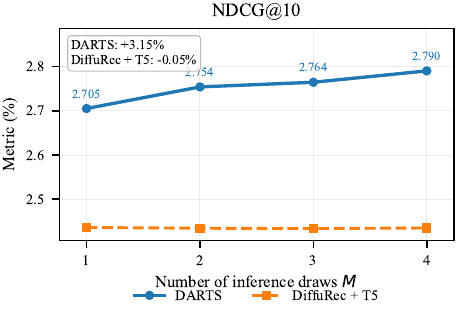}
  \caption{Posterior-predictive aggregation on \textsc{Scientific}. Ranking by the arithmetic mean of $M$ independent draws improves our NDCG@10, whereas DiffuRec remains nearly flat. Appendix~\ref{app:inference-time} reports all four metrics.}
  \label{fig:power-mean-ndcg10}
  \vspace{-\baselineskip}
\end{wrapfigure}

%% file: content/discussion.tex
\section{Discussion}
\label{sec:discussion}

Existing diffusion recommenders are generative by construction, yet their repeated samples yield nearly identical recommendation lists.
We traced this collapse to two design choices, learning the representation jointly with the denoiser and supervising the reverse process with catalogue losses, and showed by ablation that each one sharply reduces latent variation and degrades ranking.
Removing both keeps stochastic variation through decoding while the model leads every comparison in Table~\ref{tab:main-results}, and averaging several draws further improves all four ranking metrics (Figure~\ref{fig:power-mean}), a gain that injecting noise into a discriminative model does not provide (Appendix~\ref{app:stochastic-sasrec}).

\paragraph{Limitations and future work.}
Leave-one-out evaluation observes a single next interaction, so ranking metrics cannot tell whether alternative samples are plausible futures.
Multi-future settings would allow calibration, controllable diversity, and coverage to be measured directly. Structured outputs such as semantic identifiers are a natural next target as well.

%% file: appendix/ai-use-statement.tex
\subsection*{AI use statement}
We used generative AI tools to polish the manuscript, primarily to correct English grammar and improve the clarity of the writing.

%% file: content/acknowledgment.tex
\subsection*{Acknowledgments}
This project was provided with computing HPC and storage resources by GENCI at IDRIS thanks to the grant 2026-A0171015707 on the supercomputer Jean Zay's V100 partition.

%% file: appendix/implementation-details.tex
\section{Model and Implementation Details}
\label{app:implementation}

\subsection{Dataset statistics}

Table~\ref{tab:dataset-stats} reports the statistics of the preprocessed datasets described in Section~\ref{ssec:datasets}.

\input{tables/dataset-stats.tex}

\subsection{Compared models}
\label{app:baselines}

\paragraph{Sequential recommenders.}
\textbf{GRU4Rec}~\citep{hidasi2016gru4rec} models interaction sequences with gated recurrent units.
\textbf{BERT4Rec}~\citep{sun2019bert4rec} uses a bidirectional Transformer trained by masked-item prediction.
\textbf{SASRec}~\citep{kang2018sasrec} instead uses causal self-attention for next-item prediction.
\textbf{FMLP-Rec}~\citep{zhou2022fmlprec} replaces attention with MLP blocks and learnable filters that suppress noise in the frequency domain.
\textbf{HSTU}~\citep{zhai2024hstu} introduces hierarchical sequential transduction units designed to scale to large recommendation systems.
\textbf{FDSA}~\citep{zhang2019fdsa} models feature and item-level transitions with separate self-attention streams.
\textbf{S\textsuperscript{3}-Rec}~\citep{zhou2020s3rec} pretrains a sequential encoder with self-supervised objectives over items, attributes, and subsequences.

\paragraph{Generative retrieval models.}
\textbf{TIGER}~\citep{rajput2023tiger} quantises item-text embeddings into semantic identifiers with an RQ-VAE and autoregressively generates the identifier of the next item.
\textbf{LETTER}~\citep{wang2024letter} augments this tokenisation process with collaborative signals and diversity regularisation.
\textbf{ActionPiece}~\citep{hou2025actionpiece} learns context-dependent action tokens from co-occurrence patterns and uses permutation regularisation during generation.

\paragraph{Controlled baselines.}
\textbf{SASRec\,+\,T5} is our SASRec implementation with the same hybrid semantic--collaborative item representations as DARTS.
\textbf{DreamRec}~\citep{yang2023dreamrec} generates a target-item representation by conditional diffusion and uses classifier-free guidance.
\textbf{DiffuRec}~\citep{li2023diffurec} corrupts the target-item embedding and trains a Transformer denoiser conditioned on the interaction history.
\textbf{ADRec}~\citep{chen2025adrec} initialises diffusion from a pretrained collaborative item space and then optimises the sequential recommendation objective.
For the three diffusion methods, the \emph{+T5} variants replace their item tables with the same hybrid representations used by DARTS.

\paragraph{Result provenance.}
Results for GRU4Rec, BERT4Rec, SASRec, FMLP-Rec, HSTU, FDSA, S\textsuperscript{3}-Rec, TIGER, and LETTER are taken from \citet{zheng2025mtgrec}.
ActionPiece results are taken from \citet{zhong2025pctx}.
Both sources use the dataset preprocessing and evaluation protocol described in Section~\ref{ssec:datasets}.
We train SASRec\,+\,T5, DreamRec\,+\,T5, DiffuRec\,+\,T5, and ADRec\,+\,T5 ourselves in a shared codebase.

\subsection{Module block diagrams}
\label{app:blocks}

Figure~\ref{fig:blocks} details one layer of the denoiser $g_\theta$ and of the decoder $D_\psi$ introduced in Section~\ref{ssec:setup}.
In both modules, the conditioning latents enter only as the keys and values of a cross-attention and are never added to the residual stream.
The encoder $E_\phi$ is a stack of pre-norm bidirectional self-attention blocks with rotary position embeddings~\citep{su2024roformer}.

\paragraph{Denoiser.}
Each block of $g_\theta$ is an adaLN-Zero block~\citep{peebles2023dit} in which $\vz^t$ cross-attends prefix states $\vc$ (Equation~\ref{eq:diffusion-condition}).

\paragraph{Decoder.}
Each block of $D_\psi$ applies causal self-attention with rotary position embeddings to the right-shifted history item representations $\ve_i$, then cross-attends to a latent sequence of length $L$.
This sequence is $E_\phi(\tilde{s})$ in Stage~1, and $[\vc, \hat{\vz}^0]$ in Stage~3 and at inference.
The output at slot $L$ is scored against the tied item matrix, which gives the catalogue logits $\vl$.

\input{figures/blocks.tex}

\subsection{Training stages}
\label{app:training-stages}

Table~\ref{tab:training-stages} summarises the parameters trained and frozen at each of the three stages of Section~\ref{sec:method}, together with their objectives.

\input{tables/training-stages.tex}

\subsection{Inference}
\label{app:inference}

At inference, we encode $s^{\mathrm{mask}}$ once to obtain $\vc$. We sample a final latent by running the $T$ reverse steps from Gaussian noise, then pass $[\vc,\hat{\vz}^0]$ to the decoder. The final logits rank the full catalogue. Our default evaluation uses one sample. Section~\ref{ssec:generativeness} studies repeated samples that share the same conditioning prefix and differ only in their diffusion noise.

\subsection{Hyperparameters}

Table~\ref{tab:hyperparameters} lists the hyperparameters of the three training stages of Section~\ref{sec:method}.
Histories are truncated to the $20$ most recent interactions and left-padded to that length; the catalogue is extended with a padding symbol and a $\texttt{[MASK]}$ symbol, whose logits are set to $-\infty$ before ranking.
The frozen semantic table $\ve^{\text{sem}}$ is obtained by encoding each item's title and description with the \href{https://huggingface.co/sentence-transformers/sentence-t5-base}{\texttt{sentence-t5-base}} checkpoint~\citep{ni2022sentencet5} and projecting the result to $d_v$ dimensions.
All stages use a linear learning-rate schedule with warmup and gradient clipping.

Because each stage reloads the frozen components of its predecessor, the architectural settings in the upper block of Table~\ref{tab:hyperparameters} are necessarily shared across stages.
Stage~1 selects checkpoints on masked-item accuracy, Stage~2 on validation latent RMSE, and Stage~3 on NDCG@10.
Stage~2 is the longest stage in wall-clock terms despite its simple objective, because every validation pass runs the full $T$-step reverse chain; Stage~3 runs that chain on every training batch as well, which is why it uses far fewer epochs.

\input{tables/hyperparameters.tex}

%% file: tables/dataset-stats.tex
\begin{table}[!htbp]
    \caption{Preprocessed dataset statistics; Avg.\,len is the mean sequence length.}
    \label{tab:dataset-stats}
    \centering
    \small
    \begin{tabular}{lrrrrr}
        \toprule
        Dataset & \#Users & \#Items & \#Interactions & Sparsity & Avg.\,len \\
        \midrule
        Instrument & 57{,}439 & 24{,}587 & 511{,}836 & 99.964\% & 8.91 \\
        Scientific & 50{,}985 & 25{,}848 & 412{,}947 & 99.969\% & 8.10 \\
        Game & 94{,}762 & 25{,}612 & 814{,}586 & 99.966\% & 8.60 \\
        \bottomrule
    \end{tabular}
\end{table}

%% file: figures/blocks.tex
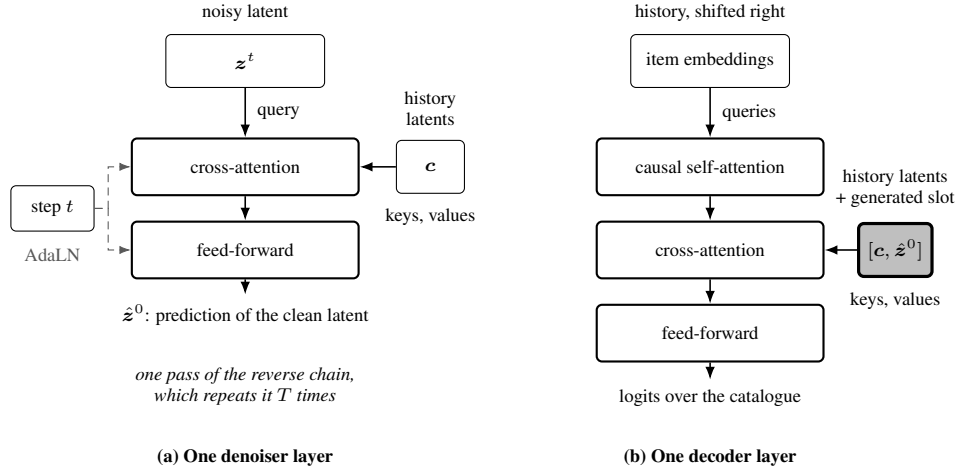
\begin{figure}[!htbp]
  \centering
  \begin{tikzpicture}[
      font=\small,
      op/.style={draw, thick, rounded corners=2pt, minimum width=30mm,
                 minimum height=7.5mm, align=center, font=\scriptsize},
      inp/.style={draw, rounded corners=2pt, minimum width=21mm, minimum height=7mm,
                  align=center, font=\scriptsize},
      key/.style={draw, rounded corners=2pt, minimum width=9mm, minimum height=7mm,
                  align=center, font=\scriptsize},
      hot/.style={key, very thick, fill=black!25},
      tstep/.style={draw, rounded corners=2pt, minimum width=11mm, minimum height=6mm,
                    align=center, font=\scriptsize},
      io/.style={font=\scriptsize, align=center},
      ar/.style={-{Latex[length=1.7mm]}, semithick},
      mod/.style={-{Latex[length=1.5mm]}, densely dashed, gray!70!black},
      ttl/.style={font=\scriptsize\bfseries, align=center},
      note/.style={font=\scriptsize\itshape, align=center},
    ]

    \begin{scope}
      \node[inp] (q) at (0,1.75) {$\vz^{t}$};
      \node[io, above=0.5mm of q] {noisy latent};

      \node[op] (ca)   at (0,0.35)  {cross-attention};
      \node[op] (ffn)  at (0,-0.75) {feed-forward};
      \node[io] (out)  at (0,-1.6)  {$\hat{\vz}^{0}$: prediction of the clean latent};

      \node[key] (mem) at (2.45,0.35) {$\vc$};
      \node[io, above=0.7mm of mem] {history\\latents};
      \node[io, below=0.7mm of mem] {keys, values};

      \draw[ar] (q) -- (ca) node[midway,right=1pt,io] {query};
      \draw[ar] (mem) -- (ca);
      \draw[ar] (ca) -- (ffn);
      \draw[ar] (ffn) -- (out);

      \node[tstep] (t) at (-2.55,-0.2) {step $t$};
      \draw[mod] (t.east) -- ++(0.18,0) |- (ca.west);
      \draw[mod] (t.east) -- ++(0.18,0) |- (ffn.west);
      \node[io, text=gray!70!black, below=1mm of t] {AdaLN};

      \node[note] at (0,-2.55) {one pass of the reverse chain,\\which repeats it $T$ times};
      \node[ttl]  at (0,-3.5)  {(a) One denoiser layer};
    \end{scope}

    \begin{scope}[xshift=6.15cm]
      \node[inp] (q2) at (0,1.75) {item embeddings};
      \node[io, above=0.5mm of q2] {history, shifted right};

      \node[op] (sa)   at (0,0.35)  {causal self-attention};
      \node[op] (ca2)  at (0,-0.75) {cross-attention};
      \node[op] (ffn2) at (0,-1.85) {feed-forward};
      \node[io] (out2) at (0,-2.7)  {logits over the catalogue};

      \node[hot] (mz) at (2.45,-0.75) {$[\vc,\hat{\vz}^{0}]$};
      \node[io, above=0.7mm of mz] {history latents\\+ generated slot};
      \node[io, below=0.7mm of mz] {keys, values};

      \draw[ar] (q2) -- (sa) node[midway,right=1pt,io] {queries};
      \draw[ar] (mz) -- (ca2);
      \draw[ar] (sa) -- (ca2);
      \draw[ar] (ca2) -- (ffn2);
      \draw[ar] (ffn2) -- (out2);

      \node[ttl] at (0,-3.5) {(b) One decoder layer};
    \end{scope}
  \end{tikzpicture}
  \caption{One layer of each module. In both, conditioning arrives from the right as the \emph{keys and values} of a cross-attention and is never added to the stream being transformed.}
  \label{fig:blocks}
\end{figure}

%% file: tables/training-stages.tex
\begin{table}[!htbp]
    \caption{The three training stages. Each stage updates only the listed parameters while the others remain frozen; the decoder $D_\psi$ is pretrained in Stage~1 and finetuned in Stage~3. Stage~2 is the only stage without a catalogue-level objective.}
    \label{tab:training-stages}
    \centering
    \small
    \setlength{\tabcolsep}{5pt}
    \begin{tabular}{llll}
        \toprule
        Stage & Trained & Frozen & Objective \\
        \midrule
        1. Autoencoder
          & $E_\phi$, $D_\psi$, $\ve^{\text{id}}$
          & $\ve^{\text{sem}}$
          & $\mathcal{L}_{\text{MLM}}$: catalogue CE on masked slots (Eq.~\ref{eq:mlm}) \\
        2. Diffusion
          & $g_\theta$
          & $E_\phi$, $D_\psi$, $\ve$
          & $\mathcal{L}_{\text{diff}}$: latent RMSE, $\vz^{0}$-prediction (Eq.~\ref{eq:diffusion}) \\
        3. Decoder
          & $D_\psi$
          & $E_\phi$, $g_\theta$, $\ve$
          & $\mathcal{L}_{\text{dec}}$: BPR on self-mined hard negatives (Eq.~\ref{eq:decoder}) \\
        \bottomrule
    \end{tabular}
\end{table}

%% file: tables/hyperparameters.tex
\begin{table}[!htbp]
    \caption{Hyperparameters for the three training stages. Architectural settings are shared across stages because later stages reload the frozen components of earlier ones. Dashes indicate settings that do not apply to a stage.}
    \label{tab:hyperparameters}
    \centering
    \small
    \begin{tabular}{llccc}
        \toprule
        & & Stage 1 & Stage 2 & Stage 3 \\
        & & Autoencoder & Diffusion & Decoder \\
        \midrule
        \multirow{7}{*}{\rotatebox{90}{Architecture}}
          & Item embedding dimension $d_v$  & \multicolumn{3}{c}{$64$} \\
          & Latent dimension $d_z$          & \multicolumn{3}{c}{$64$} \\
          & Encoder layers / heads          & \multicolumn{3}{c}{$2$ / $2$} \\
          & Decoder layers / heads          & \multicolumn{3}{c}{$2$ / $2$} \\
          & Denoiser layers / heads         & --- & \multicolumn{2}{c}{$2$ / $2$} \\
          & Dropout                         & \multicolumn{3}{c}{$0.1$} \\
          & Item mixing weight $\omega$     & \multicolumn{3}{c}{$0.5$} \\
        \midrule
        \multirow{4}{*}{\rotatebox{90}{Diffusion}}
          & Timesteps $T$                   & --- & \multicolumn{2}{c}{$200$} \\
          & Schedule                        & --- & \multicolumn{2}{c}{linear} \\
          & $\beta_1 \rightarrow \beta_T$   & --- & \multicolumn{2}{c}{$10^{-4} \rightarrow 2 \times 10^{-2}$} \\
          & Sampler                         & --- & \multicolumn{2}{c}{ancestral, $T$ steps} \\
        \midrule
        \multirow{4}{*}{\rotatebox{90}{Objective}}
          & Masking rate $p$                & $0.4$ & --- & --- \\
          & Hard negatives $n$              & --- & --- & $300$ \\
          & Temperature $\tau$              & --- & --- & $0.07$ \\
        \midrule
        \multirow{6}{*}{\rotatebox{90}{Optimisation}}
          & Optimiser                       & AdamW & Adam & AdamW \\
          & Learning rate                   & $10^{-3}$ & $10^{-4}$ & $10^{-5}$ \\
          & Weight decay                    & $10^{-3}$ & $0$ & $0$ \\
          & Batch size                      & $1024$ & $1024$ & $1024$ \\
          & Epochs                          & $600$ & $100$ & $50$ \\
          & Warmup ratio / grad.\ clip      & \multicolumn{3}{c}{$0.05$ / $1.0$} \\
        \bottomrule
    \end{tabular}
\end{table}

%% file: appendix/item-representation-ablation.tex
\section{Effect of Item Representations}
\label{app:t5-ablation}

Every model in Section~\ref{ssec:overall-performance} that we train ourselves uses hybrid semantic--collaborative item representations, in which a frozen T5 encoding of the item text is mixed with a learned identifier embedding.
Because this choice is shared across DARTS and the diffusion baselines, it cannot explain the differences between them.
It could, however, inflate the comparison against published numbers that use identifier embeddings only.
Table~\ref{tab:t5-ablation} therefore reports DARTS and each baseline on \textsc{Scientific} in both configurations.
Each pair differs by a single configuration flag, with architecture, data, optimiser, and schedule held fixed.
Our identifier-only SASRec reproduces the figure reported by~\citet{zheng2025mtgrec}, which both validates our implementation and confirms that our preprocessing and evaluation protocol match the source of the quoted baselines in Table~\ref{tab:main-results}.

\input{tables/t5-ablation.tex}

Semantic features help every model, but the size of the effect varies by more than an order of magnitude and does not track the strength of the underlying model.
DiffuRec, the strongest diffusion baseline, gains $23.2\%$ NDCG@10, while DreamRec---the weakest---gains only $10.7\%$ and remains far behind regardless of representation.
ADRec is the extreme case, improving by $190.7\%$: with identifier embeddings alone it is barely competitive (NDCG@10 $0.0067$), and semantic features recover most of the gap to the other baselines.
This is consistent with its design, which relies on a pretrained item table to warm-start the denoiser and is therefore unusually sensitive to how items are represented.

DARTS is the least sensitive of the four to this choice, gaining $14.4\%$ NDCG@10, and it is the only one whose identifier-only configuration remains competitive with the baselines in their stronger setting: at NDCG@10 $0.0236$ it improves over the best identifier-only baseline by $18.6\%$ and already matches SASRec\,+\,T5 ($0.0232$), falling only slightly short of DiffuRec\,+\,T5 ($0.0243$).

Two observations follow.
First, semantic features are not a substitute for the generative formulation: enriching DreamRec's item representations does not make it competitive, the ordering of the diffusion baselines is unchanged by the switch, and DARTS retains most of its advantage without them.
Second, the margins reported in Section~\ref{ssec:overall-performance} are measured against baselines in their stronger configuration, so they are not an artefact of the item encoder.

%% file: tables/t5-ablation.tex
\begin{table}[!htbp]
    \caption{Effect of hybrid semantic--collaborative item representations on \textsc{Scientific}. Each model is evaluated with learned identifier embeddings only and with the hybrid T5--collaborative representations used throughout Section~\ref{sec:experiments}. The last column reports the relative change in NDCG@10.}
    \label{tab:t5-ablation}
    \centering
    \small
    \setlength{\tabcolsep}{4pt}
    \begin{tabular}{lccccccccr}
        \toprule
        & \multicolumn{4}{c}{Identifier embeddings} & \multicolumn{4}{c}{Hybrid (\,+\,T5)} & \\
        \cmidrule(lr){2-5} \cmidrule(lr){6-9}
        Model & R@5 & R@10 & N@5 & N@10 & R@5 & R@10 & N@5 & N@10 & $\Delta$N@10 \\
        \midrule
        SASRec   & 0.0259 & 0.0412 & 0.0150 & 0.0199 & 0.0277 & 0.0452 & 0.0177 & 0.0232 & $+16.6\%$ \\
        DreamRec & 0.0104 & 0.0176 & 0.0067 & 0.0089 & 0.0112 & 0.0196 & 0.0071 & 0.0098 & $+10.7\%$ \\
        DiffuRec & 0.0251 & 0.0393 & 0.0152 & 0.0197 & 0.0302 & 0.0468 & 0.0190 & 0.0243 & $+23.2\%$ \\
        ADRec    & 0.0077 & 0.0137 & 0.0048 & 0.0067 & 0.0236 & 0.0374 & 0.0150 & 0.0195 & $+190.7\%$ \\
        \midrule
        \textbf{DARTS} & 0.0285 & 0.0446 & 0.0185 & 0.0236 & \textbf{0.0329} & \textbf{0.0524} & \textbf{0.0207} & \textbf{0.0270} & $+14.4\%$ \\
        \bottomrule
    \end{tabular}
\end{table}

%% file: appendix/ce-ablation.tex
\section{Discrete Supervision Inside the Diffusion Process}
\label{app:ce-ablation}

Section~\ref{sec:introduction} attributes the collapse of existing diffusion recommenders partly to supervising the denoiser with a loss defined on decoded item logits, and Section~\ref{ssec:ablations} summarises the experiment that isolates this mechanism.
This appendix gives the setup and the full sweep.

\paragraph{Setup.}
We take a single Stage-1 checkpoint and retrain Stage~2 with the mixed objective
\begin{equation}
  \label{eq:ce-ablation}
  \mathcal{L}_\alpha(\theta)
  \;=\;
  (1 - \alpha)\, \mathcal{L}_{\text{diff}}(\theta)
  \;+\;
  \alpha \, \mathcal{L}_{\text{CE}}(\theta),
  \qquad
  \mathcal{L}_{\text{CE}}(\theta)
  \;=\;
  -\, \E_{s,\, t,\, \bm{\epsilon}} \left[ \log p_\psi\!\left(i_L \,\middle|\, D_\psi\big([\vc, \hat{\vz}^{0}]\big)_L \right) \right],
\end{equation}
where $\hat{\vz}^{0} = g_\theta(\vz^{t}, t, \vc)$ is the denoiser's current prediction, decoded by the frozen decoder at every training step.
Setting $\alpha=0$ recovers the Stage-2 objective of Section~\ref{ssec:stage2}, while $\alpha=1$ supervises the denoiser with cross-entropy alone.
Nothing else changes: same initialisation, same architecture, same schedule.
Because the encoder and decoder stay frozen throughout, $\alpha$ is the only route by which a discrete objective can reach the generative stage, which is what makes the comparison clean.

\input{tables/ce-ablation.tex}

\paragraph{Results.}
Table~\ref{tab:ce-ablation} shows that the damage starts immediately and then compounds.
Adding a tenth of a cross-entropy term already more than halves $R$, from $0.30$ to $0.14$, and cuts $\mathrm{PR}$ from $4.7$ to $1.5$, leaving barely more than a single sampled direction.
Both keep falling as $\alpha$ grows: at $\alpha=1$ the variance ratio is $0.005$, sixty times smaller than at $\alpha=0$, and $\mathrm{PR}$ reaches $0.65$.
Equation~\ref{eq:participation-ratio} gives $\mathrm{PR}\ge1$ for any nonzero covariance.
In practice, we compute each context's ratio as $(\operatorname{tr}\mC_{\vc})^2/(\operatorname{tr}(\mC_{\vc}^2)+\epsilon)$ with $\epsilon=10^{-8}$, set it to zero when $\operatorname{tr}\mC_{\vc}\le10^{-8}$, and report the average over contexts.
Values below one therefore come from contexts whose samples have collapsed numerically: they are either counted as zero or, when $\operatorname{tr}\mC_{\vc}$ is of order $10^{-4}$ or smaller, shrunk by $\epsilon$.
An average below one thus signals widespread collapse, but its magnitude should not be read as an effective dimensionality.

Ranking, which the discrete term is intended to improve, never exceeds the $\alpha=0$ result and degrades monotonically: NDCG@5 is $27\%$ lower already at $\alpha=0.1$ and $57\%$ lower at $\alpha=1$.

The decoded-set dissimilarity $J$ moves the other way, rising from $0.30$ to roughly $0.52$.
Taken alone this would read as more varied recommendations, but it coincides with collapsing latent variation and collapsing accuracy, so the varied slates are not drawn from a meaningfully wider distribution: once the denoiser is trained through decoded logits, the decoder turns near-identical latents into different low-quality lists.
$J$ therefore reflects the sensitivity of the decoder as much as the breadth of the sampling, which is why we read it alongside $R$, $\mathrm{PR}$, and NDCG rather than on its own.

Cross-entropy inside the diffusion process thus costs generative behaviour and ranking accuracy at the same time, which is the recommendation analogue of the argument \citet{hu2026elf} make for language.

%% file: tables/ce-ablation.tex
\begin{table}[!htbp]
    \caption{Adding catalogue cross-entropy to the Stage-2 objective on \textsc{Scientific}. Runs differ only in the mixing weight $\alpha$ of Equation~\ref{eq:ce-ablation}: $\alpha=0$ is our latent-only objective, while $\alpha=1$ uses cross-entropy alone. Diagnostics are those of Section~\ref{ssec:generativeness}.}
    \label{tab:ce-ablation}
    \centering
    \small
    \begin{tabular}{ccccc}
        \toprule
        $\alpha$ & $J$ & $R$ $\uparrow$ & $\mathrm{PR}$ $\uparrow$ & NDCG@5 \\
        \midrule
        $\mathbf{0.0}$ & $0.30$ & $\mathbf{0.300}$ & $\mathbf{4.70}$ & $\mathbf{0.0207}$ \\
        $0.1$ & $0.43$ & $0.140$ & $1.50$ & $0.0151$ \\
        $0.3$ & $0.46$ & $0.040$ & $1.45$ & $0.0111$ \\
        $0.5$ & $0.52$ & $0.018$ & $1.02$ & $0.0102$ \\
        $0.7$ & $0.50$ & $0.010$ & $0.85$ & $0.0098$ \\
        $1.0$ & $0.52$ & $0.005$ & $0.65$ & $0.0089$ \\
        \bottomrule
    \end{tabular}
\end{table}

%% file: appendix/design-ablation.tex
\section{Effect of Updating Item Representations During Diffusion Training}
\label{app:design-ablation}

Appendix~\ref{app:ce-ablation} studies the Stage-2 objective while keeping the item representations fixed.
We now examine the second design choice: whether these representations should remain fixed or be updated during diffusion training.

\paragraph{Setup.}
Starting from the same Stage-1 checkpoint, we retrain Stage~2 under four configurations.
We either freeze the item representations, as in Section~\ref{ssec:stage2}, or update them jointly with the denoiser.
For each setting, we train with either latent regression or catalogue cross-entropy on decoded logits, corresponding to $\alpha=0$ and $\alpha=1$ in Equation~\ref{eq:ce-ablation}.
All other components and hyperparameters are unchanged.
The two frozen configurations are also reported in Table~\ref{tab:ce-ablation}.

\input{tables/design-ablation.tex}

\paragraph{Results.}
Table~\ref{tab:design-ablation} shows that keeping the representations fixed is important even when no cross-entropy is used.
Under latent regression, updating the representations reduces $R$ from $0.300$ to $0.024$, reduces $\mathrm{PR}$ from $4.70$ to $1.30$, and lowers NDCG@5 by $42\%$.
Changing only the objective to cross-entropy causes an even larger reduction when the representations remain frozen.
The combination used by DARTS, frozen representations with latent regression, achieves the highest $R$, $\mathrm{PR}$, and NDCG@5.

The effects are not additive.
When the representations are updated under cross-entropy, $R=0.050$ and $\mathrm{PR}=2.40$, both higher than in the frozen cross-entropy configuration, although NDCG@5 remains $48\%$ below DARTS.
The experiment therefore establishes the effect of each design choice separately, but does not support an ordering among the three alternative configurations.

As in Appendix~\ref{app:ce-ablation}, $J$ does not track recommendation quality.
It is lowest for the updated-representation regression run ($0.25$), whose NDCG@5 is $42\%$ below ours, and highest for the two cross-entropy runs.
We therefore interpret $J$ only together with the latent diagnostics and NDCG.

%% file: tables/design-ablation.tex
\begin{table}[!htbp]
    \caption{The two design choices, ablated jointly on \textsc{Scientific}. \emph{Item representations} are either frozen after Stage~1 or updated during Stage~2 alongside the denoiser; the Stage-2 objective is either latent regression alone or cross-entropy on decoded logits ($\alpha=1$ in Equation~\ref{eq:ce-ablation}). The first row is DARTS. Diagnostics are those of Section~\ref{ssec:generativeness}.}
    \label{tab:design-ablation}
    \centering
    \small
    \begin{tabular}{llcccc}
        \toprule
        Item representations & Stage-2 objective & $J$ & $R$ $\uparrow$ & $\mathrm{PR}$ $\uparrow$ & NDCG@5 \\
        \midrule
        \textbf{Frozen}  & \textbf{Latent regression} & $0.30$ & $\mathbf{0.300}$ & $\mathbf{4.70}$ & $\mathbf{0.0207}$ \\
        Frozen  & Cross-entropy      & $0.52$ & $0.005$ & $0.65$ & $0.0089$ \\
        Trained & Latent regression  & $0.25$ & $0.024$ & $1.30$ & $0.0121$ \\
        Trained & Cross-entropy      & $0.43$ & $0.050$ & $2.40$ & $0.0107$ \\
        \bottomrule
    \end{tabular}
\end{table}

%% file: appendix/stochastic-sasrec.tex
\section{Stochastic Discriminative Baseline}
\label{app:stochastic-sasrec}

A discriminative recommender already defines a categorical next-item distribution, even though standard inference sorts it into a deterministic ranking.
We test whether randomising this distribution provides the same benefit as sampling learned latent predictive states.
For SASRec's catalogue distribution $p_{\mathrm{SAS}}(i\mid\vc)$, draw independent Gumbel variables $g_i^{(m)}$ and form
\[
q^{(m)}(i\mid\vc)
=
\operatorname{softmax}_{i}\!\left(
\log p_{\mathrm{SAS}}(i\mid\vc)+g_i^{(m)}
\right).
\]
Sorting each perturbed vector gives a Plackett--Luce ranking; in particular, its first item is an exact categorical draw from $p_{\mathrm{SAS}}$.
More generally, given $M$ predictive distributions $q^{(m)}(i\mid\vc)$, we aggregate them with the generalised power mean
\begin{equation}
  \label{eq:power-mean}
  \hat{q}_{M,p}(i\mid\vc)
  =
  \left(
    \frac{1}{M}\sum_{m=1}^{M}q^{(m)}(i\mid\vc)^p
  \right)^{1/p},
  \qquad p\neq0.
\end{equation}
The limits $p\to0$ and $p\to\infty$ give the geometric mean and max pooling, respectively, while $p=1$ gives the arithmetic mean used for the posterior-predictive experiment in Figure~\ref{fig:power-mean}.
For this baseline, we use the checkpoint's training temperature $0.07$ and apply Equation~\ref{eq:power-mean} to the perturbed SASRec distributions.
This is a deliberately simple stochastic discriminative baseline: its variation is injected independently at the catalogue scores rather than learned as a distribution over predictive states.

\input{tables/stochastic-sasrec.tex}
\input{figures/sasrec-gumbel-sampling.tex}

Table~\ref{tab:stochastic-sasrec} shows that, for SASRec, output perturbation is not sufficient for useful stochastic recommendation.
A single Gumbel ranking loses about $80\%$ NDCG@5 and $79\%$ NDCG@10 relative to sorting SASRec's distribution.
Arithmetic aggregation ($p=1$), which matches the posterior-predictive experiment in Figure~\ref{fig:power-mean}, recovers little by $M=5$ and remains about $78\%$ and $75\%$ below deterministic SASRec on NDCG@5 and NDCG@10.
Larger powers perform similarly.

The geometric mean ($p=0$) recovers much faster, but this does not indicate complementary predictions.
Gumbel noise is additive in log-probability space, so geometric aggregation averages away the independent perturbations and converges back to the original SASRec ordering.
At $M=5$ it still remains about $28\%$ below deterministic SASRec on NDCG@5 and $26\%$ below it on NDCG@10.
Figure~\ref{fig:stochastic-sasrec} extends this sweep to $M=15$: all four metrics saturate well before that point and stay $11$--$14\%$ below the deterministic ranking, so further draws would not close the gap.
Thus, the upward geometric-mean curve reflects removal of noise introduced at $M=1$, whereas DARTS's latent samples improve beyond its own single-draw ranking.
This negative control shows that stochasticity and aggregation alone do not explain our gains; they require draws with complementary, context-dependent predictive signal.

%% file: tables/stochastic-sasrec.tex
\begin{table}[!htbp]
    \caption{Post-hoc stochastic inference for SASRec\,+\,T5 on \textsc{Scientific}. Gumbel noise is applied independently to every catalogue log-probability at temperature $0.07$. We aggregate $M$ perturbed distributions with power $p$; all powers coincide at $M=1$. Deterministic SASRec sorts the unperturbed distribution.}
    \label{tab:stochastic-sasrec}
    \centering
    \small
    \begin{tabular}{lccrrrr}
        \toprule
        Inference & $M$ & $p$ & R@5 & R@10 & N@5 & N@10 \\
        \midrule
        Deterministic & -- & -- & 0.0277 & 0.0452 & 0.0177 & 0.0232 \\
        \midrule
        Gumbel & 1 & --       & 0.0058 & 0.0104 & 0.0035 & 0.0050 \\
        Gumbel & 5 & $0$      & 0.0203 & 0.0343 & 0.0128 & 0.0173 \\
        Gumbel & 5 & $1$      & 0.0063 & 0.0123 & 0.0039 & 0.0058 \\
        Gumbel & 5 & $2$      & 0.0056 & 0.0109 & 0.0035 & 0.0051 \\
        Gumbel & 5 & $\infty$ & 0.0055 & 0.0106 & 0.0034 & 0.0050 \\
        \bottomrule
    \end{tabular}
\end{table}

%% file: figures/sasrec-gumbel-sampling.tex
\begin{figure}[!htbp]
  \centering
  \includegraphics[width=\textwidth]{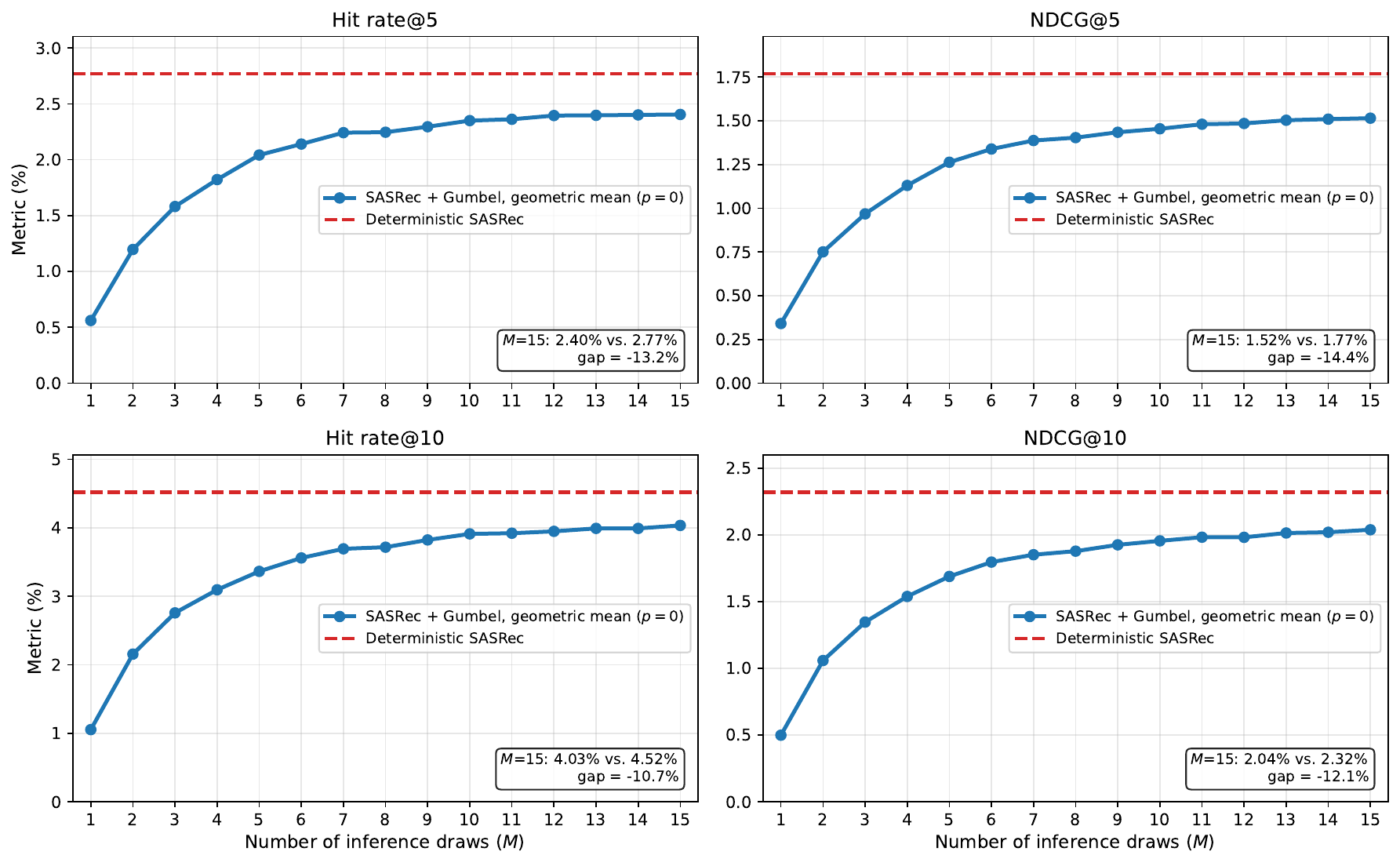}
  \caption{Stochastic SASRec on \textsc{Scientific}: geometric-mean aggregation ($p=0$) of $M$ Gumbel-perturbed catalogue distributions at a fixed top-$k$ cutoff. The dashed line is deterministic SASRec, which sorts the unperturbed distribution. Each panel is one metric; the inset boxes report the value at $M=15$ and its gap to the deterministic ranking. Additional draws only recover part of the accuracy lost at $M=1$ and saturate strictly below the deterministic ranking.}
  \label{fig:stochastic-sasrec}
\end{figure}

%% file: appendix/union-recall.tex
\section{Pooling Draws: UnionRecall}
\label{app:union-recall}

Section~\ref{ssec:ablations} spends $M$ inference draws by averaging their item distributions and ranking once on the aggregate.
The more obvious alternative is to spend them on the slate itself: run the model $M$ times, take the top-$n$ of each run, and show the user the union.
This appendix reports that variant, both because it is the natural way to read a diverse sampler and because it does not work, which is what makes the aggregation result of Section~\ref{ssec:ablations} non-trivial.

\paragraph{Definition.}
For a context $\vc$ with relevant set $G_{\vc}$, let $L^{(1)}_{\vc}, \dots, L^{(M)}_{\vc}$ be $M$ independently sampled ranked lists and let
\begin{equation}
  \label{eq:union-recall}
  U_{\vc} \;=\; \bigcup_{m=1}^{M} \operatorname{top-}n\big(L^{(m)}_{\vc}\big),
  \qquad
  \textrm{UnionRecall@}n(M) \;=\; \E_{\vc} \left[ \frac{\big| U_{\vc} \cap G_{\vc} \big|}{\big| G_{\vc} \big|} \right].
\end{equation}
Under the leave-one-out protocol of Section~\ref{ssec:datasets} we have $|G_{\vc}| = 1$, so the metric is the fraction of contexts for which \emph{at least one} of the $M$ draws places the held-out item in its top-$n$.
At $M = 1$ it is the ordinary hit rate@$n$, and it is non-decreasing in $M$ by construction.
We report the mean pooled slate size $|U_{\vc}|$ alongside it, which is the price paid: a deterministic model keeps $|U_{\vc}| = n$ for every $M$, so any growth is a direct readout of sampling diversity.

\paragraph{Results.}
Table~\ref{tab:union-recall} gives both quantities on \textsc{Scientific} for DARTS.
UnionRecall@5 rises from $0.0344$ at $M = 1$ to $0.0547$ at $M = 30$, a $59\%$ relative increase, and the slate grows from $5$ to $11.5$ items.
The union saturates well below $M \cdot n$: thirty lists of five items contain $11.5$ distinct items between them rather than $150$, so the draws are diverse but concentrated on a stable neighbourhood of the catalogue rather than wandering.
The same shape holds at $n = 10$.
Read as a diversity diagnostic this is informative, and it is the item-space counterpart of the latent measurements in Section~\ref{ssec:generativeness}: only a model that genuinely samples can grow $|U_{\vc}|$ at all.

\input{tables/union-recall.tex}

\paragraph{Why this is not an accuracy gain.}
It is tempting to compare UnionRecall@$5(M)$ against hit rate@$5$ and report the difference as an improvement, but the two are not measured at the same budget: pooling $M$ draws shows the user $|U_{\vc}| \ge 5$ items.
The fair comparison holds the number of items shown fixed, which means comparing UnionRecall@$n(M)$ at mean slate size $|U_{\vc}|$ against the single-draw hit rate@$k$ at $k = |U_{\vc}|$.
Figure~\ref{fig:union-recall} plots both curves on that common axis.

The single-draw curve dominates throughout.
At $M = 5$, pooling the top-$5$ of five draws shows $8.0$ items on average and recovers the target for $4.47\%$ of users, while simply extending one draw to its top-$8$ shows the same number of items and recovers $4.62\%$.
The gap is small but consistent, and it has the same sign at $n = 10$.
Diversity, in other words, is not by itself a source of accuracy: a longer slate from one draw uses the budget slightly better than a pooled slate from several.

\input{figures/union-recall.tex}

This is why Section~\ref{ssec:ablations} aggregates in probability space rather than in item space.
Averaging the per-draw distributions and then truncating uses the $M$ draws to sharpen the estimate of $p(i \mid \vc)$ while leaving the slate at $k$ items. The comparison against the $M = 1$ model is therefore budget-neutral, and the extra draws improve performance by roughly $3\%$ (Figure~\ref{fig:power-mean}).
The distinction matters for how the claim should be read: the value of a genuinely generative recommender here is a better posterior estimate at fixed slate size, not a longer list.

%% file: tables/union-recall.tex
\begin{table}[!htbp]
    \caption{UnionRecall@$n(M)$ and mean pooled slate size $|U_c|$ on \textsc{Scientific}, as a function of the number of independent inference draws $M$. $M = 1$ recovers the ordinary hit rate@$n$. A model whose draws were identical would keep $|U_c| = n$ and a flat recall for every $M$.}
    \label{tab:union-recall}
    \centering
    \small
    \setlength{\tabcolsep}{3.5pt}
    \begin{tabular}{lcccccccccc}
        \toprule
        $M$ & $1$ & $2$ & $3$ & $4$ & $5$ & $10$ & $15$ & $20$ & $25$ & $30$ \\
        \midrule
        UnionRecall@5  & $0.0344$ & $0.0383$ & $0.0411$ & $0.0434$ & $0.0447$ & $0.0490$ & $0.0514$ & $0.0530$ & $0.0539$ & $0.0547$ \\
        mean $|U_c|$   & $5.00$ & $6.29$ & $7.06$ & $7.62$ & $8.04$ & $9.38$ & $10.16$ & $10.72$ & $11.15$ & $11.49$ \\
        \midrule
        UnionRecall@10 & $0.0545$ & $0.0596$ & $0.0636$ & $0.0660$ & $0.0681$ & $0.0731$ & $0.0759$ & $0.0777$ & $0.0791$ & $0.0803$ \\
        mean $|U_c|$   & $10.00$ & $12.31$ & $13.65$ & $14.61$ & $15.36$ & $17.64$ & $18.95$ & $19.87$ & $20.59$ & $21.17$ \\
        \bottomrule
    \end{tabular}
\end{table}

%% file: figures/union-recall.tex
\begin{figure}[!htbp]
  \centering
  \includegraphics[width=\textwidth]{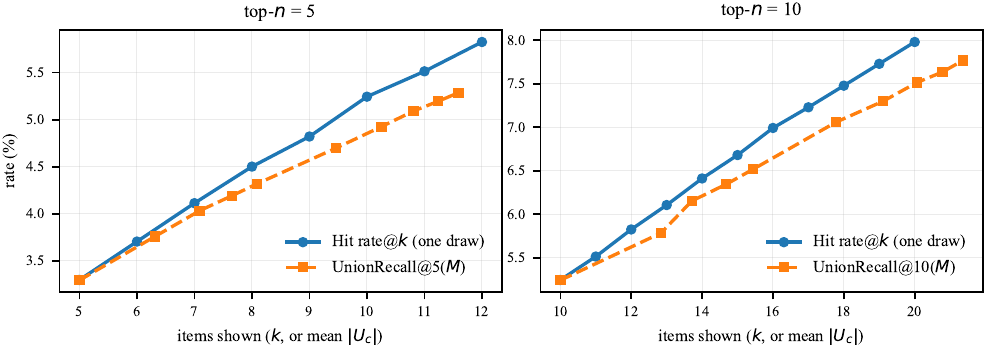}
  \caption{UnionRecall@$n(M)$ vs single-draw hit rate@$k$ on \textsc{Scientific}, both against the number of items shown (the UnionRecall curve sweeps $M$ from $1$ to $30$, each point at its mean slate size $|U_c|$). At matched slate size a single longer list matches or beats pooling several draws: sampler diversity is not by itself a source of accuracy.}
  \label{fig:union-recall}
\end{figure}

%% file: appendix/inference-time.tex
\section{Inference Cost}
\label{app:inference-time}

Section~\ref{ssec:ablations} shows that aggregating multiple independently sampled predictions can further improve ranking performance. This comes at an additional inference cost, since each draw requires a complete reverse diffusion process. We therefore measure the serving cost of DARTS and the two baselines that use the same item representations.

\input{figures/power-mean-full.tex}

\paragraph{Protocol.}

We report throughput as the number of complete predictions processed per second. A prediction includes history encoding, reverse diffusion, decoding of the final position, and catalogue scoring. All models use a latent width of $64$, a catalogue of $25{,}848$ items, and left-padded histories of length $50$. Measurements are performed at batch size $1024$ on a single NVIDIA V100 (32\,GB).

DARTS uses $200$ reverse diffusion steps and DiffuRec uses $32$ steps (as stated in the paper), matching the configurations used for the results in Table~\ref{tab:main-results}. For $M>1$, one complete prediction comprises $M$ independently sampled reverse chains and their aggregation using Equation~\ref{eq:power-mean} with $p=1$.

\input{figures/inference-throughput.tex}

\paragraph{Comparison with DiffuRec.}

At a single draw ($M=1$), DARTS achieves a throughput of $2{,}102$ predictions per second, compared with $590$ for DiffuRec+T5, while achieving an $11.1\%$ higher NDCG@10 as reported in Table~\ref{tab:main-results}. Thus, relative to DiffuRec+T5, DARTS is both faster and more accurate at the single-sample operating point.

This result is due to the substantially lower cost of each reverse step in our denoiser. DARTS cross-attends from a single latent query token to the encoded history, resulting in a per-step latency of $2.4$\,ms at this batch size. In contrast, DiffuRec re-encodes the complete item sequence with a four-block transformer at every diffusion step, requiring $54$\,ms per step. The lower per-step cost therefore more than compensates for the $6.25\times$ larger number of reverse steps in DARTS.

\paragraph{Cost of additional samples.}

Increasing $M$ requires repeating the complete reverse diffusion chain and therefore reduces throughput approximately proportionally to the number of sampled trajectories. The right panel of Figure~\ref{fig:inference-throughput} compares the cost of aggregation with the $M=1$ DiffuRec+T5 baseline. We use this reference because the samples produced by DiffuRec show substantially less variation and aggregation provides little additional ranking benefit (Figure~\ref{fig:power-mean}).

For DARTS, increasing $M$ from $1$ to $4$ improves NDCG@10 by $3.2\%$ but reduces throughput from $2{,}102$ to $629$ samples per second, corresponding to a $3.3\times$ reduction. At $M=4$, its throughput remains slightly above the $590$ samples per second of DiffuRec+T5; at $M=5$, it falls below this value.

These results characterize $M$ as an explicit quality--compute trade-off rather than a requirement of the proposed approach. In particular, the diversity reported in Section~\ref{ssec:generativeness} is obtained from a single sample ($M=1$); additional draws can be used when the incremental ranking improvement justifies the additional serving cost.

\paragraph{Comparison with a non-generative recommender.}

SASRec+T5 requires only a single forward pass and reaches $30{,}570$ samples per second, which is $14.5\times$ the throughput of DARTS at $M=1$. This higher throughput comes with a $14.2\%$ lower NDCG@10 in the reported comparison.

Thus, iterative generation introduces a substantial serving-cost premium relative to a single-pass sequential recommender. This comparison provides the computational context for the proposed generative approach: the additional inference cost enables stochastic sampling and aggregation, while the comparison with DiffuRec shows that the proposed latent denoiser can substantially reduce the per-step cost of diffusion-based recommendation.

%% file: figures/power-mean-full.tex
\begin{figure}[!htbp]
  \centering
  \includegraphics[width=\textwidth]{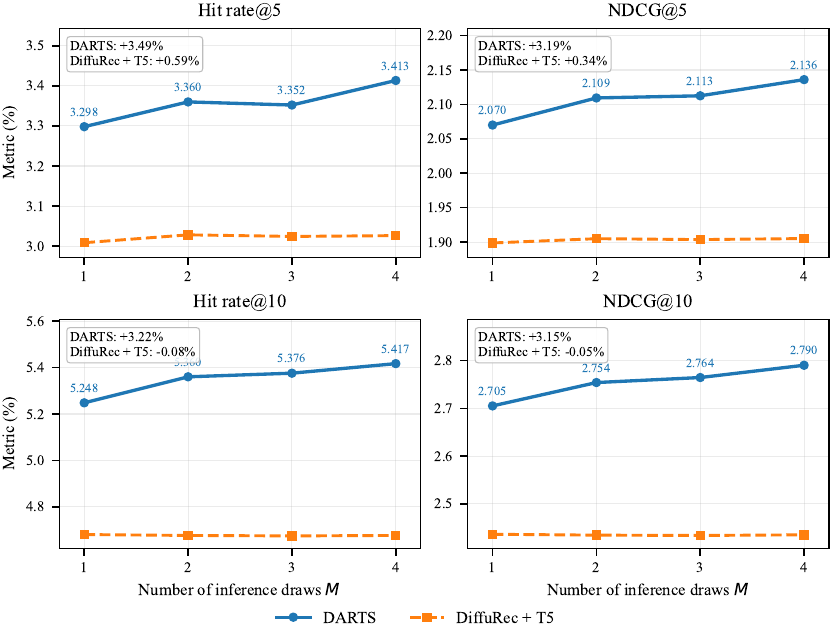}
  \caption{Posterior-predictive aggregation on \textsc{Scientific}: ranking by the arithmetic mean ($p=1$) of $M$ independent draws at a fixed top-$k$ cutoff. Each panel is one metric; the inset boxes report the relative change from $M=1$ to $M=4$ and the point labels give DARTS values in percent. DARTS turns extra draws into accuracy on every metric; DiffuRec, whose draws nearly coincide, does not.}
  \label{fig:power-mean}
\end{figure}

%% file: figures/inference-throughput.tex
\begin{figure}[!htbp]
  \centering
  \includegraphics[width=\textwidth]{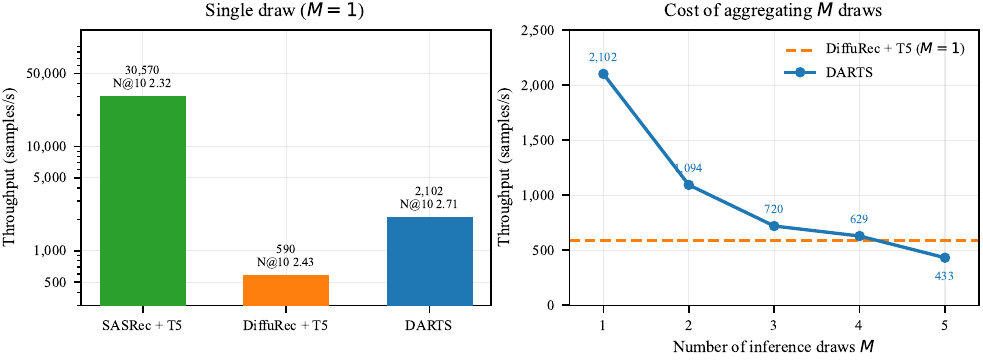}
  \caption{End-to-end inference throughput on \textsc{Scientific} at batch size $1024$ on a single NVIDIA V100 ($32$,GB). \emph{Left:} single-draw ($M=1$) throughput on a log scale, with NDCG@10 annotated. \emph{Right:} throughput as the number of aggregated draws $M$ increases. DARTS is faster than DiffuRec at $M=1$ despite using more reverse diffusion steps, while additional draws incur an approximately linear serving cost.}
  \label{fig:inference-throughput}
\end{figure}